\documentclass[12pt]{article}
\usepackage[latin1]{inputenc}
\usepackage{amsfonts,amssymb,amsmath, epsfig}

\def\Box{\leavevmode\vbox{\hrule
     \hbox{\vrule\kern4pt\vbox{\kern4pt}%
           \vrule}\hrule}}
\def\blackbox{\leavevmode\vrule height 5pt width 4pt depth 0pt\relax}
\def\endproof{\null\hfill {$\blackbox$}\bigskip}

\newcounter{appendix}
\def\appendix{\advance\c@appendix by 1
   \def\thesection{\Alph{section}}
   \ifnum\c@appendix=1 \setcounter{section}{-1} \fi
   \@startsection {section}{1}{\z@}{-3.5ex plus -1ex minus 
   -.2ex}{2.3ex plus .2ex}{\Large\bf}}

\def\paragraph#1{{\bf #1\ }}

\newtheorem{lemma}{Lemma}[section]  

\newtheorem{theorem}[lemma]{Theorem}

\newtheorem{proposition}[lemma]{Proposition}

\newtheorem{hypothesis}{Hypothesis}[section]

\title{Large scale dynamics of the Persistent Turning Walker model of fish behavior} 
\author{P. Degond $^{(1)}$, S. Motsch$^{(1)}$} 
\date{} 
\begin{document}

\maketitle

\vspace{0.5 cm}

\begin{center}
(1)\, Institute of Mathematics of Toulouse 
UMR 5219 (CNRS-UPS-INSA-UT1-UT2),
Universit\'e Paul Sabatier,
118, route de Narbonne,  31062 Toulouse cedex, 
France \\
email: degond@mip.ups-tlse.fr, motsch@mip.ups-tlse.fr
\end{center}

\vspace{0.5 cm}
\begin{abstract}
This paper considers a new model of individual displacement, based on fish motion, the so-called Persistent Turning Walker (PTW) model, which involves an Ornstein-Uhlenbeck process on the curvature of the particle trajectory. The goal is to show that its large time and space scale dynamics is of diffusive type, and to provide an analytic expression of the diffusion coefficient. Two methods are investigated. In the first one, we compute the large time asymptotics of the variance of the individual stochastic trajectories. The second method is based on a diffusion approximation of the kinetic formulation of these stochastic trajectories. The kinetic model is a Fokker-Planck type equation posed in an extended phase-space involving the curvature among the kinetic variables.  
We show that both methods lead to the same value of the diffusion constant. We present some numerical simulations to illustrate the theoretical results.

\end{abstract}

\medskip
\noindent
{\bf Acknowledgements:} The authors wish to thank Guy Théraulaz and Jacques Gautrais of the 'Centre de Recherches sur la Cognition Animale' in Toulouse,  for introducing them to the model and for stimulating discussions. 


\vskip 0.4cm


\setcounter{equation}{0}
\section{Introduction}
\label{sec_intro}

This paper considers a new model of individual displacement, the so-called 'Persistent Turning Walker' (PTW) model, which has recently been introduced to describe fish behavior \cite{Gautrais_Theraulaz}. The fish evolves with a velocity of constant magnitude and its trajectory is subject to random turns (i.e. random changes of curvature) on the one hand and to curvature relaxation to zero on the other hand. The random changes of curvature can be interpreted as a way for the fish to explore its surroundings while relaxation to zero curvature just expresses that the fish cannot sustain too strongly curved trajectories and when the curvature becomes too large, it gets back to a straight line trajectory. The combination of these two antagonist behaviors gives rise to an Ornstein-Uhlenbeck process on the curvature. The curvature is the time derivative of the director of the velocity, while the velocity itself is the time derivative of position. The PTW
process collects all these considerations into a system of stochastic differential equations. 

This model is, to the knowledge of the authors, original, and has appeared for the first time in the works by Gautrais, Theraulaz, and coworkers \cite{Gautrais_Theraulaz}. The present paper considers the large time and space scale dynamics of a two-dimensional particle subject to this PTW process. It rigorously shows (in the mathematical sense) that, at large scales, the dynamics of the particle can be described by a diffusion process and it provides a formula for the diffusion coefficient. To prove this result, two methods are considered. 

In the first method, the stochastic differential system itself is considered and the variance of the position is shown to behave, at large times, like a linear function of time. The diffusion coefficient is identified as the slope of this linear function. Because the curvature and the velocity angle can be explicitly computed, an explicit formula for the diffusion coefficient, involving some special functions, can be obtained. 

The second method considers the forward Kolmogorov equation of the stochastic process. This equation gives the evolution of the probability distribution function of the particle in the extended phase space (position, velocity angle, curvature) as a function of time. It is a Fokker-Planck type equation. The passage from the microscopic  to the macroscopic scales relies on a rescaling of the Kolmogorov equation. This rescaling depends on a small parameter $\varepsilon \ll 1$, which describes the ratio of the typical microscopic to macroscopic space units. After this rescaling, the problem has the typical form of the diffusion approximation of a kinetic problem (see references below). The goal is then to study the behaviour of the solution as $\varepsilon \to 0$. It is shown that the solution converges to some 'thermodynamical equilibrium' which is a Gaussian distribution of the curvature and a uniform distribution of the velocity angle. The equilibrium depends parametrically on the density which satisfies a spatial diffusion equation.  

Finally, the connection between the two methods is made by showing that the diffusion tensor in the second approach can be represented by a formula involving the solution of the stochastic differential equation of the first approach. Additionally, this representation leads to explicit computations which show that the two formulas for the diffusion coefficient actually coincide. This seemingly innocuous result is actually quite powerful. Indeed, the diffusion approximation method leads to a non-explicit expression of the diffusion coefficient, involving the moments of a particular solution of a stationary equation involving the leading order operator of the Fokker-Planck equation. That this non-explicit formula is equivalent to the explicit formula given by the stochastic trajectory method is by far not obvious. In this respect, the stochastic trajectory method is more powerful than the diffusion approximation approach, because it directly leads to the most simple expression of the diffusion constant. 

A third route could have been taken and has been dismissed. This third method would actually use the stochastic differential equation itself to perform the diffusion approximation in the forward Kolmogorov equation. We have preferred to use partial differential equation techniques. One reason for this choice is that these techniques can be more easily extended to more complex situations. One typical example of these more complex situations are the nonlinear systems which are obtained when interactions between individual are included.  The inclusion of interactions between individuals within the PTW model is actually work in progress. 

From the biological viewpoint, one should not restrict the content of the paper to the sole expression of the diffusion coefficient. Indeed, once interactions between individuals will be included in the PTW model, it is not clear at all that the explicit computations which led to this expression will still be tractable. In the absence of an explicit solution of the stochastic differential system, there is little grasp to get information about the large scale behaviour of the system. By contrast, the diffusion approximation approach gives a systematic tool to study the large scale behavior of such systems, in all kinds of situations, be they linear or nonlinear. By its flexibility and its versatility, the diffusion approximation approach is the method of choice to study these problems. 

One of the most popular models to describe fish behavior is the discrete Couzin-Vicsek algorithm (CVA) \cite{Aldana_Huepe,couzin02:_collec_memor,Gregoire_Chate,Vicsek95:_novel} (see also \cite{aoki82,brillinger02:_employ,orsogna06:_self,mogilner03:_mutual, parrish03:_traff,parrish02:_self} for related  models). 
For a large scale modeling of fish behavior, it is efficient to look at continuum models, which use macroscopic variables such as mean density, mean velocity and so on. Several such models based on phenomenological observations, exist (see e.g. \cite{edelstein-keshet01:_mathem,mogilner99,topaz04:_swarm,Topaz_Bertozzi_Lewis}). Several attempts to derive continuum models from the CVA model are also reported in the literature \cite{KRZB,RBKZ,RKZB}. In \cite{Degond_Motsch_CV,Degond_Motsch_CV_CRAS}, a derivation of a continuum model from a kinetic version of the CVA model is proposed. However, few Individual Based Models for fish have been validated against experimental data with a comparable care as in \cite{Gautrais_Theraulaz} for the PTW process. As such, the continuum model derived in this paper has a firm experimental basis, although further work needs certainly to be done to fully validate its biological foundations. Additional references on swarm aggregation and fish schooling can be found in \cite{camazine02:_self_organ_biolog_system}. Among other types of animal societies, ants have been the subject of numerous studies and the reader can refer (among others) to \cite{jost07:_from, theraulaz02:_spatial}, and references therein. 

The derivation of macroscopic models from particle behavior has been initiated by the seminal works of Boltzmann, and later Hilbert, Chapman and Enskog. We refer to \cite{CIP} for a mathematical perspective and to  \cite{degond03:_macros_boltz} for
an introduction to the subject from a modeling perspective.  More recently, the derivation of macroscopic models from microscopic behavior has been very productive 
in other context like traffic \cite{aw02:_deriv,helbing01:_traff} or supply-chains
\cite{armbruster06}. Diffusion
approximation problems for kinetic equations have been widely studied in the literature, in the context of neutron transport (see e.g.  \cite{BSS,BLP}), semiconductors \cite{75,72,GP,Poupaud1}, plasmas \cite{70,63,Degond_Gallic_TTSP} or polymeric fluids \cite{71}.

The outline of the paper is as follows : in section \ref{sec_PTW}, the PTW process is introduced and the main results are stated. In section \ref{sec_trajectory}
the diffusion coefficient is obtained by direct inspection of the trajectories of the stochastic differential system. In section \ref{sec_asymptotic}, the diffusion approximation of the forward Kolmogorov equation of the stochastic process is performed. Section \ref{sec_equivalence} is devoted to proving that the trajectory method and the diffusion approximation method give rise to the same value of the diffusion coefficient. In section \ref{sec_numeric}, the theoretical results are illustrated by and complemented with some numerical simulations. A conclusion is drawn in section \ref{sec_conclu}. Several proofs of auxiliary results, which are inessential for the main discussion are collected in three appendices (A, B and C).

\setcounter{equation}{0}
\section{The Persistent Turning Walker model: presentation and main results}
\label{sec_PTW}

\begin{figure}[h]
  \centering
  \includegraphics[scale=.5]{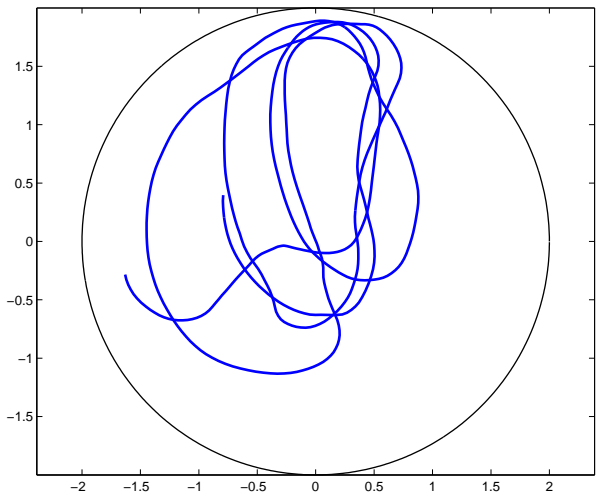}
  \caption{One trajectory of a Kuhlia Mugil fish}
  \label{fig:exp}
\end{figure}

The starting point of the present work is a new model of fish motion based on experimental data taken from experiments run in La Réunion islands during years 2001
and 2002 \cite{Gautrais_Theraulaz}. The studied species is a pelagic fish named Kuhlia Mugil. Its typical size ranges between 20 and 25 cm. The first experiments have been made with a single fish in a basin of 4 meters diameter during two minutes. A video records the positions of the fish every 12-th of a second (see figure \ref{fig:exp}). Then the data have been statistically analyzed and a model has been extracted \cite{Gautrais_Theraulaz}. 

The conclusion of the statistical analysis is that the trajectories are well described by the following systems of stochastic differential equations: 
\begin{eqnarray}
\frac{d\vec{x}}{dt} &=& c \vec{\tau}(\theta), \label{pos_unscaled} \\
\frac{d\theta}{dt} &=& c \kappa , \label{angle_unscaled}\\
d \kappa &=& -a \kappa\,dt + b\, dB_t , \label{curv_unscaled}
\end{eqnarray}
where $\vec{x}=(x_1,x_2) \in {\mathbb R}^2$ is the (two-dimensional) position vector of the (centroid of the) fish,
$\vec{\tau}(\theta)=(\cos \theta\, , \, \sin \theta)$ is the director of the velocity vector with the angle 
$\theta \in \Pi={\mathbb R}/2\pi{\mathbb Z}$ measured from the $x_1$ direction, $\kappa \in {\mathbb R}$ is the curvature of the trajectory and $dB_t$ is the standard Brownian motion. The magnitude of the velocity is constant and denoted by $c>0$. The constant $a$ is a relaxation frequency and $b$ quantifies the intensity of the random curvature jumps. $b$ has the dimension of $1/(L\sqrt{T})$ where $L$ and $T$ stand for the dimensions of length and time. 

The $\kappa$-dynamics is a standard Ornstein-Uhlenbeck process. The term ``$b \, dB_t$'' models a diffusion process in curvature space while the term ``$-a \, K\,dt$'' expresses the tendency of the individual to return to a straight line trajectory. The curvature cannot increase endlessly as a consequence of the diffusion process, but rather, must relax to zero and the relaxation is stronger as the curvature gets larger. This model has been called the \emph{Persistent Turning Walker} model (PTW) because it allows large excursions of the curvature towards positive or negative values, during which the spinning of the trajectory persists for a certain time.

We stress the difference with more standard diffusion processes (such as those suffered by photons in a diffusive medium), in which the Brownian motion acts on the velocity itself (or, in the case of a velocity of constant magnitude, on the angle $\theta$). In this case, the diffusion process acts on the second derivative of the particle positions, and the associated kinetic equation is of Fokker-Planck type. This model of photon diffusion is also relevant for a certain number of animal species \cite{Othmer_Hillen}. 

In the PTW model, the diffusion process acts on the curvature, i.e. on the third derivative of the position vector. An intuitive justification of the relevance of this model for animal behaviour is by considering the non-differentiability of the Brownian motion. Because of this feature, the photon diffusion process involves infinite second derivatives of the position, i.e. infinite forces. However, an animal body can only exert finite forces and the muscles act only in such a way that the velocity angle undergoes smooth variations. The PTW model precisely presents this feature of having smooth second order derivatives, i.e. smooth forces.

Our goal in the present work is to study the large-scale dynamics of the stochastic differential system (\ref{pos_unscaled})-(\ref{curv_unscaled}). This is best done in scaled variables, where the dimensionless parameters of the model are highlighted. We use $t_0 = a^{-1}$ as time unit, $x_0 = c a^{-1}$ as space unit, and $\kappa_0 = x_0^{-1}$ as curvature unit, and we introduce the dimensionless time, space and curvature as $t' = t/ t_0$, $x' = x/ x_0$ and $\kappa' = \kappa/ \kappa_0$. For simplicity, we omit the primes. In scaled variables, the PTW model is written: 
\begin{eqnarray}
\frac{d\vec{x}}{dt} &=& \vec{\tau}(\theta), \label{pos_scaled} \\
\frac{d\theta}{dt} &=&  \kappa , \label{angle_scaled}\\
d \kappa &=& -\kappa\,dt + \sqrt 2 \alpha dB_t , \label{curv_scaled}
\end{eqnarray}
where the only dimensionless parameter left is $\alpha$ such that 
\begin{eqnarray}
\alpha^2 = \frac{b^2 c^2}{2a^3} , \label{alpha_def}
\end{eqnarray}
The meaning of $\alpha^2$ is the following: $b/\sqrt a$ is the amplitude of a curvature change during a relaxation time $a^{-1}$, while $c/a$ is obviously the distance travelled by the particle during this time. The product of these two quantities is dimensionless and is equal to $\sqrt 2  \alpha$. It quantifies the strength of the curvature jumps relative to the other phenomena.

The individual dynamics can be translated in terms of a probability distribution $f(t,\vec{x},\theta,\kappa) \, d\vec{x} \, d\theta \, d\kappa$ of finding particles 
at times $t$ with position in small neighborhoods $ d\vec{x} \, d\theta \, d\kappa$ of position $\vec{x}$, velocity angle $\theta$ and curvature $\kappa$. The link between the individual dynamics and the evolution of the probability distribution $f$ is given by the forward Kolmogorov equation :
\begin{equation}
  \label{eq:kolmo}
  \partial_t f + \vec{\tau}\cdot \nabla_{\vec{x}} f  +  \kappa \partial_\theta f -
  \partial_\kappa (\kappa f) - \alpha^2 \partial_{\kappa^2} f = 0.
\end{equation}
This equation is an exact transcription of the individual dynamics, where the initial value $f_0$ at time $t=0$ is given by the probability distribution of the initial conditions of the stochastic differential system (\ref{pos_scaled})-(\ref{curv_scaled}). For more detailed considerations about the  forward Kolmogorov equation and its link with stochastic differential systems, we refer the reader to \cite{oksendal92:_stoch,bass97:_diffusb}.

In order to capture the macroscopic dynamics, two possible routes can be taken, using either the stochastic differential system (\ref{pos_scaled})-(\ref{curv_scaled}) or the partial differential equation (\ref{eq:kolmo}). In this work, we follow both routes and verify that they lead to the same large-scale behaviour. The advantage of working directly on the stochastic system is that it is simpler and it leads to explicit formulas. However, as soon as the system gets more complicated, and in particular nonlinear, explicit solutions can no longer be found and this methodology can hardly be pursued. On the other hand, the PDE approach, which, in the present case is more complicated, is also more systematic and more general. In particular, it is generally usable in the more complex nonlinear cases (see e.g. \cite{Degond_Motsch_CV,Degond_Motsch_CV_CRAS}). A particular important complex situation is the case of many interacting fish. In future work, we plan to extend the PTW model to populations of interacting fish and to use the PDE approach to extract the large-scale dynamics of the system.

From the analysis of the individual trajectories, 
explicit exact expressions for $\kappa$ and $\theta$ in terms of stochastic
integrals can be found. Unfortunately, there is no such explicit result for the position $\vec{x}(t)$, but we can
calculate the first two moments of the probability distribution of $\vec{x}(t)$ explicitly,  using the expressions of $\kappa$ and $\theta$. We show that the mean of the position vector stays at the origin: ${\mathbb E}\{\vec{x}(t)\} = (0,0)$ (where ${\mathbb E}$ denotes the expectation over all sources of randomness, in the initial data and in the stochastic process) and that the variance grows asymptotically  linearly in time. More exactly, we prove:

\begin{theorem}
  \label{thm:asymp}
Under assumptions on the initial conditions that will be specified later on (see (\ref{pos_ini})-(\ref{indep_ini})), the solution of system (\ref{pos_scaled})-(\ref{curv_scaled}) satisfies: 
\begin{equation}
\mathrm{Var}\{\vec{x}(t)\} \stackrel{t\rightarrow +\infty}{\sim} 2\mathcal{D}\,t, \quad \mbox{with } \quad \mathcal{D} = \int_0^\infty \exp \left(-\alpha^2(-1+s+{\mathrm e}^{-s}) \right)\,ds.     \label{eq:D}
\end{equation}
\end{theorem}

The notation Var is for the variance over all sources of randomness. The asymptotic linear growth of the variance (\ref{eq:D}) suggests that the dynamics of the system is of  diffusive type at large times with diffusion coefficient $\mathcal{D}$.
We can find an expression of $\mathcal{D}$ in terms of special functions. Indeed, we have
\begin{proposition}
The following expression holds true:
\begin{eqnarray}
& & \mathcal{D} = \left( \frac{{\mathrm e}}{\alpha^2} \right)^{\alpha^2} \gamma(\alpha^2,\alpha^2), 
\label{eq:D_2}
\end{eqnarray}
where $\gamma(z,u)$ is the incomplete gamma function: 
\begin{eqnarray}
& & \gamma(z,u) = \int_0^u  {\mathrm e}^{-t} \, t^{z-1} \, dt .  
\label{incomplete_gamma}
\end{eqnarray}
$\mathcal{D}$ has the following series representation: 
\begin{eqnarray}
& & \mathcal{D} = {\mathrm e}^{\alpha^2} \sum_{n=0}^{\infty} \frac{(-1)^n \alpha^{2n}}{n! \, (n+ \alpha^2)} . 
\label{eq:D_3}
\end{eqnarray}
\label{prop_gamma}
\end{proposition}

To investigate the large scale dynamics of the solution of the kinetic equation (\ref{eq:kolmo}) (the existence of which can be easily proved, see proposition \ref{ppo:exis}), we need to rescale the variables to the macroscopic scale. 
Indeed, in eq. (\ref{eq:kolmo}), all the coefficients are supposed to be of order unity. This means that the time and space scales of the experiment are of the same order as the typical time and length scales involved in the dynamics, such as, the relaxation time or the inverse of the typical random curvature excursions. Of course, in most experiments, this is not true, since the duration of the experiment and the size of the experimental region are large compared with the time and length scales involved in the dynamics. 

To translate this observation, we change the space unit $x_0$ to a new space space unit $x'_0 = x_0/\varepsilon$, where $\varepsilon \ll 1$ is a small parameter. This induces a change of variables $x' = \varepsilon x$. We make a similar operation on the time unit $t'_0 = t_0/\eta$, $t' = \eta t$ with $\eta \ll 1$. Now, the question of linking $\eta$ to $\varepsilon$ is a subtle one and is largely determined by the nature of the asymptotic regime which is achieved by the system. In the present case, we expect that the asymptotic regime will be of diffusive nature, in view of 
theorem \ref{thm:asymp} and so, we will investigate the so-called 'diffusion approximation' which involves a quadratic relationship between $\eta$ and $\varepsilon$: $\eta = \varepsilon^2$. 

For this reason, we introduce the diffusive rescaling: 
\begin{eqnarray}
  t'=\varepsilon^2 t \quad ; \quad \vec{x}'=\varepsilon \vec{x}, \label{rescaling}
\end{eqnarray}
and we make the following change of variable in the distribution $f$ :
\begin{displaymath}
  f^{\varepsilon}(t',\vec{x}',\theta,\kappa) = \frac{1}{\varepsilon^2}\,f\left( \frac{t'}{\varepsilon^2}, \frac{\vec{x}'}{\varepsilon},\theta,\kappa\right).
\end{displaymath}
The scaling of the magnitude of the distribution function is unnecessary, since the problem is linear. However, it is chosen in order to preserve the total number of particles. 
Introducing (\ref{rescaling}) into (\ref{eq:kolmo}) leads to the following problem for $f^\varepsilon$:
\begin{equation}
  \label{eq:fep}
  \varepsilon \partial_t f^\varepsilon + \vec{\tau}\cdot \nabla_{\vec{x}} f^\varepsilon  +  \frac{1}{\varepsilon}\,[ \, \kappa \partial_\theta f^\varepsilon -
  \partial_\kappa (\kappa f^\varepsilon) - \alpha^2 \partial_{\kappa^2} f^\varepsilon \, ] = 0
\end{equation}
In order to analyze the large-scale dynamics of (\ref{eq:fep}), we need to investigate the limit $\varepsilon \rightarrow 0$. We show that $f^\varepsilon$ converges to an equilibrium distribution function (i.e. a function which cancels the $O(\varepsilon^{-1})$ term of (\ref{eq:fep})) $f^0$ which depends parametrically on the particle density $n^0(x,t)$ and $n^0$ evolves according to a diffusion equation. More precisely, we prove: 

\begin{theorem}
\label{thm:fasymp}
Under hypothesis \ref{eq:hyp} on the initial data to be precised below, the solution $f^\varepsilon$ of (\ref{eq:fep}) converge weakly in a Banach space also to be specified below, (see (\ref{eq:X:LM2})) $X$:
\begin{eqnarray}
f^\varepsilon \stackrel{\varepsilon \rightarrow 0}{\rightharpoonup } n^0\,\frac{M(\kappa)}{2\pi} \quad
\text{ in } X \, \mbox{ weak star} , \label{feps_to_f0}
\end{eqnarray}
where $M$ is a Gaussian distribution of the curvature with zero mean and variance $\alpha^2$ (see \ref{eq:M}) and $n^0= n^0(x,t)$ is the solution  of the system:
\begin{eqnarray}
& & \partial_t n^0 + \nabla_{\vec{x}} \cdot J^0 = 0 , 
\label{continuity} \\
& & J^0 = - D \nabla_{\vec{x}} n^0 , \label{current} 
\end{eqnarray}
where the initial datum $n_0^0$ and the diffusion tensor $D$ will be defined later on (see (\ref{ini_cond}) and (\ref{eq:tenseur}) respectively). 
\end{theorem}

The following theorem connects the two methods by showing that the tensor $D$ is related to ${\mathcal D}$ given by (\ref{eq:D}): 

\begin{theorem}
\label{thm:Dexplicit}
The tensor $D$ defined by (\ref{eq:tenseur}) satisfies :
\begin{eqnarray}
    D = \frac{\mathcal{D}}{2}\,\mbox{Id}, \label{D=D}
\end{eqnarray}
where ${\mathcal D}$ is given by (\ref{eq:D}) and Id denotes the $2 \times 2$ identity tensor. 
\end{theorem}

This theorem confirms that the trajectory method and the asymptotic PDE method are equivalent. The factor $2$ between the two coefficients comes from the dimension of the problem. Indeed, ${\mathcal D}$ is the variance of $|x|^2= |x_1|^2+|x_2|^2$ while $D$ is the diffusion coefficient in a given direction.

\setcounter{equation}{0}
\section{Large-scale dynamics of the PTW model by the trajectory method}
\label{sec_trajectory}

In this section, we want to show theorem \ref{thm:asymp} and proposition \ref{prop_gamma}. We first specify the initial
conditions. First, we fix the starting point of the particle at the origin :
\begin{eqnarray}
  \vec{x}(t=0) = (0,0). \label{pos_ini}
\end{eqnarray}
We suppose that the initial velocity angle is uniformly distributed on
the one-dimensional sphere, i.e. :
\begin{eqnarray}
  dP \{\theta|_{t=0} = \theta \} = \frac{d \theta}{2 \pi}. \label{angle_ini}
\end{eqnarray}
For the curvature, we make the following observation: eq. (\ref{curv_scaled}) predicts that the process $\kappa(t)$ converges
exponentially fast to its stationary state, which is a Gaussian distribution with zero mean and variance equal to $\alpha^2$ \cite{oksendal92:_stoch}.  We denote such a Gaussian distribution by $\mathcal{N}(0,\alpha^2)$. For this reason, we suppose: 
\begin{eqnarray}
  dP \{ \kappa|_{t=0} = \kappa \} =  \mathcal{N}(0,\alpha^2)(\kappa). \label{curv_ini}
\end{eqnarray}
The last hypothesis on the initial conditions is the following:
\begin{eqnarray}
\mbox{The processes } 
\, \theta(t=0),\, \kappa(t=0) \text{ and } B_t \text{ are independents}. 
\label{indep_ini}
\end{eqnarray}

We stress that this choice of initial conditions is for simplicity only. Completely arbitrary initial conditions would lead to the same large time behaviour, but the computation would be slightly more complicated. Since we are mainly interested in the explicit expression of ${\mathcal D}$, a choice of initial conditions which simplifies the calculations is legitimate.

We begin by proving the following proposition: 

\begin{proposition}
\label{ppo:moments}
The solution of the stochastic differential equation (\ref{pos_scaled})-(\ref{curv_scaled}) with initial
condition given by (\ref{pos_ini})-(\ref{indep_ini}) satisfies :
\begin{eqnarray}
\label{eq:mean}
{\mathbb E}\{\vec{x}(t)\} &=& (0,0) \; , \qquad \forall\,t\geq 0, \\
\label{eq:var}
\mathrm{Var}\{\vec{x}(t)\} &=& 2 \!\! \int_{s=0}^{t} \!\!(t-s) \exp\left(-\alpha^2\left(- 1+s + {\mathrm e}^{-s}\right)\right)\,ds.
\end{eqnarray}
\end{proposition}

To prove this proposition, we first establish explicit formulae for the solutions of (\ref{angle_scaled}) and (\ref{curv_scaled}). The proof is deferred to appendix A. 
\medskip
  
\begin{lemma}
\label{lem:eds}
The solution of the stochastic differential system  (\ref{angle_scaled}), (\ref{curv_scaled}) with initial conditions (\ref{angle_ini})-(\ref{indep_ini})
is given by: 
\begin{eqnarray}
\theta(t) &=& \theta_0 + \kappa_0 -\kappa(t) + \sqrt{2}\alpha B_t , \label{angle_explicit} \\
\kappa(t) &=& {\mathrm e}^{-t} \kappa_0 + \sqrt{2}\alpha {\mathrm e}^{-t} \int_0^t {\mathrm e}^{s}\,dB_s . \label{curv_explicit}
\end{eqnarray}
Additionally, 
\begin{eqnarray}
\theta(t) = \theta_0 + K_0^t,  \label{angle_explicit_2}
\end{eqnarray}
where $K_0^t$ is a Gaussian random variable independent of $\theta_0$ with zero mean and variance $\beta_t^2$ given by :
\begin{eqnarray}
\beta_t^2 = \mathrm{Var}\{K_0^t \} = 2 \alpha^2(-1+t+{\mathrm e}^{-t}). \label{beta_explicit}
\end{eqnarray}
\end{lemma}
\medskip

\noindent
{\bf Proof of proposition \ref{ppo:moments}:}  Using Lemma \ref{lem:eds}, we can compute the first two moments of $\vec{x}(t)$. Let us start
with the computation of the mean. If we write $\vec{x}(t) = (x_1(t)\,,\,x_2(t))$, we have :
\begin{displaymath}
x_1(t) = \int_0^t \cos \theta(s) \,ds \quad , \quad x_2(t) = \int_0^t \sin \theta(s) \,ds,
\end{displaymath}
and, computing the mean :
\begin{displaymath}
{\mathbb E}\{x_1(t)\} = {\mathbb E}\left\{ \int_0^t \cos \theta(s) \,ds \right\} = \int_0^t {\mathbb E}\left\{\cos \theta(s)\right\}\,ds.
\end{displaymath}
Now,  we can develop $\theta(s)$ using (\ref{angle_explicit_2}):
\begin{displaymath}
{\mathbb E}\left\{\cos \theta(s) \right\} = {\mathbb E}\left\{\cos(\theta_0 + K_0^s)\right\} = {\mathbb E}\left\{\cos \theta_0 \,  \cos K_0^s - \sin \theta_0 \, \sin K_0^s \right\}.
\end{displaymath}
By the independence of $\theta_0$ and $K_0^s$ we finally have:
\begin{displaymath}
{\mathbb E}\left\{\cos \theta(s)\right\} = {\mathbb E} \{\cos \theta_0\} {\mathbb E}\{\cos K_0^s \} -
{\mathbb E}\{\sin(\theta_0)\}  {\mathbb E}\{\sin K_0^s \} =0,
\end{displaymath}
since the expectations of $\cos \theta_0$ and $\sin \theta_0$ over the uniform probability distribution on $\theta_0$ are zero. Finally, we have $ {\mathbb E}\{x_1(t)\} = 0$, and similarly for $x_2$. This proves (\ref{eq:mean}).

Now for the variance of $\vec{x}(t)$, we write:
\begin{eqnarray}
    \mathrm{Var}\{\vec{x}(t)\} = {\mathbb E}\{x_1^2(t)+x_2^2(t)\} = 2 {\mathbb E}\{x_1^2(t)\}. \label{var_x}
\end{eqnarray}
by the isotropy of the problem. Then, 
\begin{eqnarray*}
    {\mathbb E}\{x_1^2(t)\} &=& {\mathbb E}\left\{\left(\int_0^t \cos \theta(s)\,ds\right)^2\right\} \\
    &=& \int_0^t \int_0^t {\mathbb E}\{\cos \theta(s) \,  \cos \theta(u) \}\,dsdu \\
    &=& 2 \int_0^t du \int_0^u ds \, \,  {\mathbb E}\{\cos \theta(s) \cos \theta(u)\}.
\end{eqnarray*}
Since $u \geq s$, we can write $\theta(u)$ as follows :
  \begin{displaymath}
    \theta(u) = \theta_0 + \int_0^s \kappa(z)\,dz + \int_s^u \kappa(z)\,dz = \theta_0
    + K_0^s + K_s^u
  \end{displaymath}
where $K_0^s$ and $K_s^u$ are Gaussian random variables independent of $\theta_0$ with zero mean and variances $\beta_s^2$ and $\beta_{u-s}^2$ respectively, thanks to (\ref{beta_explicit}). Then, using standard identities for trigonometric functions, we get
\begin{eqnarray*}
    {\mathbb E}\{\cos \theta(s) \,  \cos \theta(u) \} &=& {\mathbb E}\{\cos(\theta_0 +K_0^s)  \cos(\theta_0+K_0^s +
    K_s^u)\} \\
    &=& \frac{1}{2}\,(\cos(2\theta_0+ 2K_0^s + K_s^u) + \cos(-K_s^u)).
\end{eqnarray*}
But since $\theta_0$ is independent of $K_0^s$ and $K_s^u$ we have ${\mathbb E}\{\cos(2\theta_0 + 2K_0^s + K_s^u)\} =0$ since the mean of a $\cos (\theta_0 + C)$ over the uniform distribution of $\theta_0$ is zero whatever the value of $C$.
Then :
  \begin{eqnarray*}
    {\mathbb E}\{\cos(\theta(s)) \cos(\theta(u))\} &=& \frac{1}{2}\,{\mathbb E}\{\cos\left(-K_s^u\right)\}\\
    &=& \frac{1}{2}\,\int_{{\mathbb R}} \cos(y)\frac{1}{\sqrt{2\pi}\beta_{u-s}}
    \,{\mathrm e}^{-\frac{y^2}{2\beta_{u-s}^2}}\,dy \\
    &=&
    \frac{1}{2}\,{\mathrm e}^{-\frac{1}{2}\beta_{u-s}^2}.  
  \end{eqnarray*}
Indeed, an elementary computation shows that for any  Gaussian random variable $Z$ with zero mean and variance $\sigma^2$, one has 
\begin{eqnarray}
 {\mathbb E}\{ \cos(Z) \} = \exp(-\sigma^2/2). 
 \label{cos_Z})
\end{eqnarray}
Thus, 
  \begin{eqnarray*}
    {\mathbb E}\{x_1(t)^2\} &=& \int_{u=0}^t \int_{s=0}^u \exp\left(-\alpha^2\left(-1+|u-s| + {\mathrm e}^{-|u-s|}\right)\right)\,dsdu.
  \end{eqnarray*}
  Using the change of unknowns $w=u-s$ and $y=u$ and inverting the order of integration we find :
  \begin{displaymath}
    {\mathbb E}\{x_1(t)^2\} = \int_{w=0}^t (t-w)\exp\left(-\alpha^2\left(- 1+w + {\mathrm e}^{-w}\right)\right)\,dw.
  \end{displaymath}
Using (\ref{var_x}), we finally find (\ref{eq:var}), which ends the proof of the proposition. \endproof

In order to prove \ref{thm:asymp}, we investigate the behavior of the variance Var$\{\vec x(t) \}$  (given by (\ref{eq:var})) when $t \rightarrow +\infty$.

\medskip

\noindent
{\bf End of proof of Theorem \ref{thm:asymp}:} we write, thanks to (\ref{eq:var}): 
  \begin{displaymath}
    \mathrm{Var}\{\vec{x}(t)\} - 2\mathcal{D}t = -2 \int_{s=0}^t s {\mathrm e}^{-\alpha^2(-1+s+{\mathrm e}^{-s})}\,ds - 2
    \int_{s=t}^\infty t {\mathrm e}^{-\alpha^2(-1+s+{\mathrm e}^s)}\,ds.
  \end{displaymath}
  We have to show that the difference is bounded independently of $t$. For the first term, we have:
  \begin{displaymath}
    \left| \int_{s=0}^t s {\mathrm e}^{-\alpha^2(-1+s+{\mathrm e}^{-s})}\,ds \right| \leq
    \int_0^t {\mathrm e}^{\alpha^2} s{\mathrm e}^{-\alpha^2 s}\,ds,
  \end{displaymath}
and integrating by parts, we find :
  \begin{displaymath}
    \left| \int_{s=0}^t s {\mathrm e}^{-\alpha^2(-1+s+{\mathrm e}^s)}\,ds \right| \leq
    \frac{{\mathrm e}^{\alpha^2}}{\alpha^2} \left[ -t {\mathrm e}^{-\alpha^2 t} -
      \frac{{\mathrm e}^{-\alpha^2 t}}{\alpha^2}+\frac{1}{\alpha^2} \right] \leq C_1.
  \end{displaymath}
  For the second term, we have:
  \begin{displaymath}
    \left|\int_{s=t}^\infty t {\mathrm e}^{-\alpha^2(-1+s+{\mathrm e}^s)}\,ds \right| \leq t
    \int_t^\infty {\mathrm e}^{\alpha^2} {\mathrm e}^{-\alpha^2 s}\,ds \leq t
    {\mathrm e}^{\alpha^2} \frac{{\mathrm e}^{-t\alpha^2}}{\alpha^2} \, \leq C_2.
  \end{displaymath}
This proves that the difference is $\mathrm{Var}\{\vec{x}(t)\} - 2\mathcal{D}t$ is bounded independently of $t$ and completes the proof. \endproof

We now prove Proposition \ref{prop_gamma} which gives an explicit approximation of the diffusion coefficient. This approximation is useful for practical simulations. 

\medskip 
\noindent
{\bf Proof of Proposition \ref{prop_gamma}:} the change of variables $t = \alpha^2 {\mathrm e}^{-s}$ in the integral (\ref{eq:D}) leads to (\ref{eq:D_2}). The series representation (\ref{eq:D_3}) follows from a similar series representation of the incomplete gamma function (see e.g. formula (8.354) of \cite{GR}). The series representation can also be found by expanding the exponential in the integral (\ref{incomplete_gamma}) in power series. This point is left to the reader. \endproof

\setcounter{equation}{0}
\section{Large-scale dynamics of the PTW model through the diffusion approximation of the associated kinetic equation}
\label{sec_asymptotic}

\subsection{Formal asymptotics}
\label{sub_formal}

In this section, for the reader's convenience, we give a formal proof of theorem \ref{thm:fasymp}. We write (\ref{eq:fep}) as follows: 
\begin{equation}
  \label{eq:fep_2}
  \varepsilon \partial_t f^\varepsilon + \vec{\tau}\cdot \nabla_{\vec{x}} f^\varepsilon  +  \frac{1}{\varepsilon}\,A f^\varepsilon = 0
\end{equation}
where we define the operator $A$ acting on functions $u(\theta, \kappa)$ as follows:
\begin{equation}
\label{eq:A}
Au =  \kappa \partial_\theta u - \partial_\kappa (\kappa u) - \alpha^2
\partial_{\kappa^2} u,
\end{equation}

The formal investigation of the limit $\varepsilon \to 0$ usually starts by considering the Hilbert expansion (see e.g. \cite{degond03:_macros_boltz} for the general theory or \cite{Degond_Gallic_TTSP} for an application in the context of Fokker-Planck equations): 
\begin{equation}
\label{Hilbert}
f^\varepsilon = f^0 + \varepsilon f^1 +  O(\varepsilon^2), 
\end{equation}
with $f^k$ being independent of $\varepsilon$ and inserting it into (\ref{eq:fep_2}). Then, collecting all the the terms of comparable orders with respect to $\varepsilon$, we are led to a sequence of equations. The first one, corresponding to the leading $O(\varepsilon^{-1})$ term is $Af^0 = 0$, which means that $f^0$ lies in the kernel of $A$. In section \ref{sub_rigorous}, we show that the kernel of $A$ is composed of functions of the form $f^0(t,\vec x, \theta, \kappa) = n^0(t, \vec x) M(\kappa)/(2 \pi)$ where $M(\kappa)$ is a normalized Gaussian with zero mean and variance $\alpha^2$: 
  \begin{equation}
    \label{eq:M}
    M(\kappa) = \frac{1}{\sqrt{2\pi}\alpha^2}\,{\mathrm e}^{-\frac{\kappa^2}{2\alpha^2}},
  \end{equation}
and $n^0(t, \vec x)$ is a function still to be determined. 

In order to determine $n^0$,  we first integrate (\ref{eq:fep_2}) with respect to $(\theta, \kappa) \in \Pi \times {\mathbb R}$ and use that $\int Au \, d\theta \, d \kappa = 0$. Defining the density $n^\varepsilon(t,\vec x)$ and the flux $J^{\varepsilon}(t,\vec{x})$ by 
\begin{eqnarray}
  n^{\varepsilon}(t,\vec{x}) = \int_{\theta,\kappa} f^{\varepsilon} \, \,d\kappa \,  d\theta, \quad 
    J^{\varepsilon}(t,\vec{x}) = \int_{\theta,\kappa} \frac{f^{\varepsilon}}{\varepsilon}\, \vec{\tau}(\theta)\,d\kappa d\theta.
    \label{neps_jeps}
\end{eqnarray}
we find: 
  \begin{eqnarray}
  \partial_t n^{\varepsilon} + \nabla_{\vec x} \cdot J^{\varepsilon} = 0. 
    \label{continuity_eps}
  \end{eqnarray}
We note that this continuity equation is valid for all values of $\varepsilon$.
Then, letting $\varepsilon \to 0$, we formally have $n^{\varepsilon} \to n^0$. If we prove that $J^0$ given by (\ref{current}) is the limit of $J^{\varepsilon}$, as $\varepsilon \to 0$, then, we can pass to the limit in (\ref{continuity_eps}) and find (\ref{continuity}). 

System (\ref{continuity}) and (\ref{current}) is a diffusion system, which completely determines $n^0(t, \vec x)$, given its initial datum $n^0_0(\vec x)$. Here, for simplicity, we assume that the initial datum for (\ref{eq:fep_2}) is of the form $f^\varepsilon (0,\vec x,\theta, \kappa) = n_0(\vec x) M(\kappa)/(2 \pi)$ and the resulting initial condition for $n^0$ is therefore $n_0^0= n_0$ (in this formal convergence proof, we admit the functions and the convergences are as smooth as required). 

So, the only points left in the proof are the existence of a limit for $J^{\varepsilon}$ and the validity of (\ref{current}) for $J^0$. Note that the existence of a limit is not obvious because of the factor $\varepsilon$ at the denominator of the integral (\ref{neps_jeps}) defining $J^\varepsilon$. To prove that the limit exists, we use the Hilbert expansion (\ref{Hilbert}) again and compute $f^1$. Since $A$ is linear, collecting the terms of order $O(\varepsilon^0)$ leads to:  
  \begin{eqnarray}
  -A f^1 =  \frac{M(\kappa)}{2 \pi} \, \, \vec \tau \cdot \nabla_{\vec x} n^0 \, . 
    \label{Af1=}
  \end{eqnarray}
Again using the linearity of $A$ and the fact that it operates only with respect to the $(\theta, \kappa)$ variables, we can write the solution of (\ref{Af1=}) as 
$f^1 = - \vec \chi \cdot \nabla_{\vec x} n^0$, where $\vec \chi = (\chi_1, \chi_2)$ is a solution of the problem 
  \begin{eqnarray}
  A \vec \chi =  \frac{M(\kappa)}{2 \pi} \, \, \vec \tau  \, .
    \label{auxil}
  \end{eqnarray}
This equation must be understood componentwise (i.e $\chi_1$ is associated with $\tau_1 = \cos \theta$ and $\chi_2$ with $\tau_2 = \sin \theta$). Since the right-hand side of (\ref{auxil}) has zero average with respect to $(\theta,\kappa)$, proposition \ref{ppo:solva} below shows that it has a unique solution, up to an element of the kernel of $A$. We can single out a unique solution by requesting that $\vec \chi$ has zero average with respect to $(\theta,\kappa)$ as well. Then, all solutions $f^1$ to (\ref{Af1=}) can be written as 
  \begin{eqnarray}
  f^1 =   - \vec \chi \cdot \nabla_{\vec x} n^0  +  n^1(t, \vec x) M(\kappa)/(2 \pi)\, ,
    \label{f1=}
  \end{eqnarray}
where the second term of (\ref{f1=}) is an arbitrary element of the kernel of $A$. We shall see that the determination of $n^1$ is unnecessary. 

Now, inserting the Hilbert expansion (\ref{Hilbert}) into the integral (\ref{neps_jeps}) defining $J^\varepsilon$, we find: 
\begin{eqnarray}
J^{\varepsilon}(t,\vec{x}) &=& \frac{1}{\varepsilon} \int_{\theta,\kappa} f^0 \, \vec{\tau}(\theta)\,d\kappa d\theta + \int_{\theta,\kappa} f^1 \, \vec{\tau}(\theta)\,d\kappa d\theta + O(\varepsilon) \nonumber \\
&=& \hspace{1.55cm} 0  \hspace{1.55cm} +  \int_{\theta,\kappa} f^1 \, \vec{\tau}(\theta)\,d\kappa d\theta + O(\varepsilon) , 
    \label{jeps_expan}
\end{eqnarray}
because $f^0$ is independent of $\theta$ and $\int \vec \tau(\theta) \, d \theta = 0$. Therefore, $J^\varepsilon$ has a limit when $\varepsilon \to 0$ and this limit is given by 
\begin{eqnarray}
J^0(t,\vec{x}) =  \int_{\theta,\kappa} f^1 \, \vec{\tau}(\theta)\,d\kappa d\theta , 
    \label{j0=intf_1}
\end{eqnarray}
To compute $J^0$ we insert expression (\ref{f1=}) into (\ref{j0=intf_1}) and find
\begin{eqnarray}
J^0(t,\vec{x}) =  \int_{\theta,\kappa} (- \vec \chi \cdot \nabla_{\vec x} n^0  +  n^1 M/(2 \pi)) \, \vec{\tau}(\theta)\,d\kappa d\theta , 
    \label{j0=}
\end{eqnarray}
The second term vanishes and the first one can be written
\begin{eqnarray}
J^0(t,\vec{x}) =  - \left( \int_{\theta,\kappa} \vec \tau \otimes \vec \chi \, d \theta \, d \kappa \right) \,  \nabla_{\vec x} n^0 ,     \label{j0=2}
\end{eqnarray}
which is nothing but formula (\ref{current}) with the diffusivity tensor $D$ given by  (\ref{eq:tenseur}).

This shows the formal convergence of the solution of the Fokker-Planck equation (\ref{eq:fep}) to that of the diffusion system (\ref{continuity}), (\ref{current}). 

Now, to make this proof rigorous, we need to justify all the formal convergences. In the framework of the Hilbert expansion, this requires to work out the regularity of the various terms of the expansion. This is doable and actually leads to stronger convergences than the one we are going to prove, but this is a bit technical (see e.g. \cite{Degond_Gallic_TTSP}). 

What we are going to do instead is proving a convergence result in a weaker topology without using the Hilbert expansion technique. The method is close to the so-called moment method, which consists in integrating the equation against suitable test functions. This convergence proof is developed in section \ref{sub_rigorous}, but before that, we state an existence result for the original Fokker-Planck equation (\ref{eq:fep_2}).

\subsection{Functional setting and existence result}
\label{sub_functional}

We define the differential operator $D$ acting on smooth functions $f(\kappa)$ by :
  \begin{equation}
    \label{eq:op_D}
    Df = \partial_\kappa (\kappa f) + \alpha^2 \partial_{\kappa^2} f.
  \end{equation}
We state some properties of $D$, the proofs of which are easy and left to the reader. We recall that $M(\kappa)$ denotes the normalized Gaussian with zero mean and variance $\alpha^2$ (\ref{eq:M}). 

\begin{proposition}
 Let $f$ and $g$ be smooth functions decreasing at infinity. The following identities hold true: 
\begin{eqnarray}
& &     Df = \alpha^2\,\frac{\partial}{\partial \kappa} \left(M \, \frac{\partial}{\partial \kappa} \frac{f}{M} \right),     \label{eq:dec_D} \\
& &  \int_{{\mathbb R}} Df \, \,  g\,\frac{d\kappa}{M} = -\alpha^2 \, \int_{{\mathbb R}}  \, M \, \partial_\kappa
    \left( \frac{f}{M} \right) \partial_\kappa \left( \frac{g}{M}\right)\,d\kappa =  \int_{{\mathbb R}} f \, 
    Dg\,\frac{d\kappa}{M} , \label{D_weak} \\
& & \int_{{\mathbb R}} Df \,   f\,\frac{d\kappa}{M} = -\alpha^2 \,\int_{{\mathbb R}}  M \, \left| \partial_\kappa
    \left( \frac{f}{M} \right) \right|^2 \,d\kappa \, \leq 0, \label{D_dissip} \\
& & Df =0 \Leftrightarrow \exists c \in {\mathbb R}, \quad f= c M. \label{D_ker}
\end{eqnarray}
\label{prop_properties_D}
\end{proposition}

The first identity translates the fact that $M$ is the stationary measure of the Ornstein-Uhlenbeck process. The second one that $D$ is formally self-adjoint with respect to the measure $d \kappa /M$. The third one shows that $D$ is dissipative. The same inequality holds with any non-decreasing function $\eta(f)$, indeed, $\int D(f) \, \eta(f) \, M^{-1} \,  d \kappa \leq 0$. If $\eta$ is the logarithm function, the corresponding quantity would be the relative entropy dissipation of $f$ with respect to $M$. Entropy plays an important role in kinetic theory (see \cite{CIP} for a review). Finally, the last quantity states that the kernel of $D$ is one-dimensional and spanned by $M$. 

Proposition \ref{prop_properties_D} shows that the natural $L^2$ norm associated with this operator has a weight $M^{-1}$ and that the natural $H^1$ semi-norm is given by the right-hand side of (\ref{D_dissip}). This motivates the introduction of the following functional spaces, endowed with their naturally associated Hilbert structures and norms:
\begin{eqnarray}
\label{eq:V}
\label{eq:H}
H &=& \{ u :\, \Pi_{\theta} \times {\mathbb R}_\kappa \rightarrow {\mathbb R} \,/\; \int_{\theta,\kappa}
|u(\theta,\kappa)|^2\,\frac{d\theta d\kappa}{M} < + \infty \}, \\
V &=& \left\{ u \in H \,/ \;  \int_{\Pi,{\mathbb R}}  M  \left|\partial_\kappa
\left(\frac{u}{M}\right)\right|^2 \,d\kappa d\theta < +\infty
\right\}, \\
\label{eq:X:LM2}
L_M^2 &=& L^2({\mathbb R}_{\vec{x}}^2, H), \quad X = L^2([0,T] \times {\mathbb R}_{\vec{x}}^2, V).
\end{eqnarray}
Identifying $H$ with its dual, with have a Hilbertian triple $V \subset H \subset V'$, where $V'$ is the dual of $V$ and all injections are continuous. They are not compact because $V$ does not bring any regularity with respect to $\theta$.

The existence proof follows closely the existence proof of \cite{Degond_AnnENS} (see appendix A of this reference) and for this reason, is omitted (see also \cite{Degond_Gallic_TTSP}). The proof relies on an existence theorem due to J. L. Lions \cite{lions61:_equat}. 

\begin{proposition}
\label{ppo:exis}
Let $\varepsilon >0$. We assume that $f_0$ belongs to $L_M^2$ defined by (\ref{eq:X:LM2}). Then there exists a unique solution $f^\varepsilon$ to (\ref{eq:fep}) with initial datum $f^0$  in the class of functions $Y$ defined by :
\begin{displaymath}
Y=\left\{ f \in X \,/\; \partial_t f + \varepsilon^{-1} \vec{\tau}\cdot \nabla_{\vec{x}} f + \varepsilon^{-2} \kappa \partial_\theta f \in X' \right\}.
\end{displaymath}
Moreover, we have the inequality for any $T>0$:
\begin{equation}
\label{eq:fep_borne}
||f^\varepsilon(T)||_{L^2_M}^2 + \frac{\alpha^2}{\varepsilon^2} \int_0^T \int_{\vec x,\theta,\kappa}  M  \left|\partial_\kappa \left(\frac{f^\varepsilon}{M}\right)\right|^2 \,d\kappa \, d\theta \, d \vec x \, dt = ||f^\varepsilon(0)||_{L^2_M}^2 \, .
\end{equation}
\end{proposition}

\medskip
Estimate (\ref{eq:fep_borne}) is obtained via a Green and a trace formula for functions belonging to $Y$ which can be deduced from the one proved in \cite{Degond_AnnENS}.

\subsection{Rigorous asymptotics}
\label{sub_rigorous}

\medskip
We first study operator $A $ given by (\ref{eq:A}), i.e. $Af = \kappa\partial_\theta f - Df $ \, and state some properties which will be proved in appendix B. We view $A$ as an unbounded operator on the Hilbert space $H$ with domain $D(A)$ given by: 
\begin{displaymath}
 D(A) = \left\{ u(\theta,\kappa) \in V\;/\, Au \in H \right\}.
\end{displaymath}

\begin{lemma}
\label{lem:A_max}
Operator $A$ is maximal monotone. Moreover its kernel (or Null-space) is given by:
\begin{eqnarray}
    \mathrm{Ker}(A) = \{cM \, , \; c\in {\mathbb R} \}, \label{kerA}
\end{eqnarray}
with $M$ defined by (\ref{eq:M}).
\end{lemma}

\medskip
\begin{lemma}
\label{lem:A*}
The adjoint $A^*$ of $A$ in $H$ is given by $A^*f = - \kappa\partial_\theta f - Df $. 
It is a Maximal monotone operator with domain $D(A^*) = D(A)$ and Ker$(A^*) =$ Ker$(A)$. 
\end{lemma}

\medskip
\begin{proposition}
\label{ppo:solva}
Let $g \in H$. Then, there exists $u \in D(A)$ such that 
\begin{equation}
    \label{eq:solva}
    Au=g,
\end{equation}
if and only if $g$ satisfies the following solvability condition: 
\begin{eqnarray}
    \int_{\theta,\kappa} g(\theta,\kappa)\,d\theta d\kappa = 0. \label{solva}
\end{eqnarray}
Moreover, the solution $u$ is unique up to a constant times $M$. A unique solution can be singled out by presribing the condition 
  \begin{eqnarray}
    \int_{\theta,\kappa} u(\theta,\kappa)\,d\theta d\kappa = 0. \label{cancel}
  \end{eqnarray}
The same lemma applies to the equation $A^* u = g$.
\end{proposition}

\medskip
As an application of this lemma, let $\vec{\chi}$ be the solution of :
  \begin{eqnarray}
    A \vec{\chi} = \vec{\tau}(\theta)\,\frac{M}{2\pi}, \label{def_chi}
  \end{eqnarray}
  with $\vec{\tau}(\theta)=(\cos \theta\, , \, \sin \theta)$. Since $\tau$ has zero average over $\theta$ and $\kappa$, $\vec{\chi}$ is well-defined and unique thanks to Proposition \ref{ppo:solva}. Then, we define the tensor $D$ by:
  \begin{equation}
    \label{eq:tenseur}
    D = \int_{\theta,\kappa} \vec{\tau}(\theta) \otimes \vec{\chi} \,d\theta d\kappa.
  \end{equation}
Note that, since $\int_{\theta,\kappa}
  \vec{\tau}(\theta) M(\kappa)\,d\theta d\kappa=0$, it would not change the value of $D$ to add any element of Ker$(A)$ to $\vec{\chi}$. 
  
\begin{lemma}
Let $R$ denote the reflection operator $u(\theta, \kappa) \to Ru(\theta, \kappa) = u (\theta, -\kappa)$. Then, $\vec \chi^* = R \vec \chi$ is the unique solution (satisfying (\ref{cancel})) of 
  \begin{eqnarray}
    A^* \vec{\chi}^* = \vec{\tau}(\theta)\,\frac{M}{2\pi}, \label{chi_*}
  \end{eqnarray}
and we have
\begin{equation}
    \label{eq:tenseur_*}
    D = \int_{\theta,\kappa} \vec{\tau}(\theta) \otimes \vec{\chi}^* \,d\theta d\kappa.
\end{equation}
\label{lem_chi*}
\end{lemma}

\noindent
{\bf Proof:} Obviously, $D$ commutes with $R$: $DR = RD$ while $\kappa \partial_\theta$ anticommutes with $R$: $\kappa \partial_\theta (Ru) = - R(\kappa \partial_\theta u)$. Therefore, $RA = A^* R$. Since the right-hand side of (\ref{def_chi}) is invariant by $R$, applying $R$ to both sides of (\ref{def_chi}) leads to (\ref{chi_*}). Then, the change of variables $\kappa' = - \kappa$ in the integral at the right-hand side of (\ref{eq:tenseur_*}) shows that it is equal to $D$. \endproof

\medskip
To study the limit $\varepsilon \rightarrow 0$, we make the following hypothesis on the initial conditions.
\begin{hypothesis}
  \label{eq:hyp}
  We suppose that the initial condition $f_0^\varepsilon$ is uniformly bounded in $L_M^2$ and converges weakly in $L_M^2$ to $f_0^0$ as $\varepsilon \to 0$. 
\end{hypothesis}
We can now prove theorem \ref{thm:fasymp}. The initial datum for the diffusion system (\ref{continuity}, (\ref{current}) will be shown to be:
\begin{eqnarray}
& & n^0_{t=0} = n_0^0 = \int_{\theta,\kappa} f_0^0(\vec{x},\theta,\kappa)\,d\kappa d\theta. 
\label{ini_cond} 
\end{eqnarray}

\medskip
\noindent
{\bf Proof of theorem \ref{thm:fasymp}:} By hypothesis \ref{eq:hyp} inequality (\ref{eq:fep_borne}) implies :
\begin{equation}
\label{eq:fep_borne_2}
||f^\varepsilon(T)||_{L^2_M}^2 + \frac{\alpha^2}{\varepsilon^2} \int_0^T \int_{\vec x,\theta,\kappa}  M  \left|\partial_\kappa \left(\frac{f^\varepsilon}{M}\right)\right|^2 \,d\kappa \, d\theta \, d \vec x \, dt \leq C \, ,
\end{equation}
with $C$ independent of $\varepsilon$. So $(f^\varepsilon)_\varepsilon$ is a bounded sequence in $L^\infty(0,T, L^2_M)$ and satisfies 
\begin{equation}
\label{eq:fep_borne_3}
\int_0^T \int_{\vec x,\theta,\kappa}  M  \left|\partial_\kappa \left(\frac{f^\varepsilon}{M}\right)\right|^2 \,d\kappa \, d\theta \, d \vec x \, dt \leq C \varepsilon^2\, ,
\end{equation}
for any time interval $T$ (by the diagonal process, we will eventually be able to take an increasing sequence of times $T$ tending to infinity, so that the result will be valid on the whole interval $t \in (0,\infty)$). Therefore, there exists $f^0 \in L^\infty(0,T, L^2_M)$ and a subsequence, still denoted by $f^\varepsilon$,  satisfying :
  \begin{displaymath}
    f^\varepsilon \stackrel{\varepsilon \rightarrow 0}{\rightharpoonup } f^0 \quad
    \text{ in } L^\infty(0,T, L^2_M) \mbox{ weak star }.
  \end{displaymath}
Furthermore, with (\ref{eq:fep_borne_3}), we deduce that $f^0 = C(\vec x,\theta,t) M(\kappa)$.  Then, letting $\varepsilon \to 0$ in (\ref{eq:fep}), we get that $Af^0 = 0$ in the distributional sense. This implies that $C(\vec x,\theta,t)$ is independent of $\theta$ and we can write 
\begin{eqnarray}
f^0(t, \vec x, \theta, \kappa)=n^0(t,\vec{x})\frac{M(\kappa)}{2\pi} ,  
\label{leading_order}
\end{eqnarray}
the quantity $n^0(t,\vec{x}) = \int f^0(t, \vec x, \theta, \kappa) \, d \theta \, d \kappa$ being the density associated with $f^0$. 
  
Our next task is to show that $n^0$ satisfies the diffusion model (\ref{continuity}), (\ref{current}) with initial condition (\ref{ini_cond}). 
We first note that $f^\varepsilon$ is a week solution of  (\ref{eq:fep}) with initial condition $f_0^\varepsilon$ in the following sense: $f^\varepsilon$ satisfies:
\begin{equation}
    \label{eq:fep_weak}
    \int_0^T \int_{\vec{x},\theta,\kappa}  f^\varepsilon (-\varepsilon\partial_t \varphi - \vec{\tau}\cdot
      \nabla_{\vec{x}} \varphi+ \frac{1}{\varepsilon} A^*(\varphi))\,\frac{d\kappa
    d\theta  d\vec{x}}{M} dt = \varepsilon \int_{\vec{x},\theta,\kappa} f_{0}^\varepsilon \varphi_{t=0}\, \frac{d\kappa d\theta d\vec{x}}{M}.
  \end{equation}
for all test functions $\varphi$  in the space $C_c^2 ([0,T) \times {\mathbb R}_{\vec{x}}^2 \times \Pi_\theta \times {\mathbb R}_\kappa)$ of twice continuously differentiable functions with compact support in $[0,T) \times {\mathbb R}_{\vec{x}}^2 \times \Pi_\theta \times {\mathbb R}_\kappa$. Again, the trace at $t=0$ has a meaning, thanks to a trace formula for functions in $Y$ which is proven in \cite{Degond_AnnENS}. 

We recall the definition of the flux (\ref{neps_jeps}). We prove that $J^{\varepsilon}$ has a weak limit as $\varepsilon \to 0$. To this aim, in the weak formulation (\ref{eq:fep_weak}), we take as a test function  $\varphi=\vec \phi(t,\vec{x}) \cdot \vec{\chi}^*(\theta,\kappa)$ with  $\vec{\chi}^*$ the auxiliary function defined as the solution to (\ref{chi_*}) and $\vec \phi$ is a smooth compactly supported vector test function of $(\vec x,t)$. Although $\varphi$ does not have a compact support, a standard truncation argument (which is omitted here) can be used to bypass this restriction. This allows us to write:
  \begin{displaymath}
    \int_0^T \int_{\vec{x},\theta,\kappa}  [ f^\varepsilon (-\varepsilon\partial_t  \ - (\vec{\tau}\cdot
    \nabla_{\vec{x}})) ( \vec \phi \cdot \vec{\chi}^*) + \frac{1}{\varepsilon} f^\varepsilon \vec{\tau} \frac{M}{2\pi} \cdot \vec \phi  ] \,\frac{d\kappa
    d\theta  d\vec{x}}{M} dt = \varepsilon \int_{\vec{x},\theta,\kappa} f_{0}^\varepsilon \vec \phi_{t=0} \cdot \vec{\chi}\, \frac{d\kappa d\theta d\vec{x}}{M}.
  \end{displaymath}
Taking the limit $\varepsilon \rightarrow 0$, we find :
  \begin{eqnarray*}
    \lim_{\varepsilon \rightarrow 0} \int_0^T \int_{\vec{x}} J^{\varepsilon} \cdot  \vec \phi\,d\vec{x} dt &=& 2 \pi \, \int_0^T
    \int_{\vec{x},\theta,\kappa} f^0 (\vec{\tau}\cdot \nabla_{\vec{x}})( \vec \phi \cdot \vec{\chi}^*)\, \frac{d\kappa
      d\theta  d\vec{x}}{M} dt \\
    &=& \int_0^T \int_{\vec{x}} n^0 \, \nabla_{\vec{x}} \cdot \left( \left( \int_{\theta,\kappa}
    \vec{\chi}^* \otimes \vec{\tau} \, d\kappa  d\theta \right)^T  \vec \phi \right)\, d\vec{x} dt,
  \end{eqnarray*}
where the exponent T denotes the transpose of a matrix.   
Using (\ref{eq:tenseur_*}) and taking the limit $\varepsilon \to 0$ shows that $J^\varepsilon$ converges weakly (in the distributional sense) towards $J^0$ satisfying 
  \begin{eqnarray*}
     \int_0^T \int_{\vec{x}} J^0 \cdot \vec \phi\,d\vec{x} dt 
    &=& \int_0^T \int_{\vec{x}} n^0 \, \nabla_{\vec{x}} \cdot \left( (D)^T  \vec \phi \right)\, d\vec{x} dt. 
  \end{eqnarray*}
This last equation is the weak form of eq. (\ref{current}).

Finally, to prove (\ref{continuity}), we apply the weak formulation (\ref{eq:fep_weak}) to a test function of the
  form $\varphi=\phi(t,\vec{x})M(\kappa)$, where again, $\phi(x,t)$ is a scalar, smooth and compactly supported test function of $(\vec x,t)$ in ${\mathbb R}^2 \times [0,T)$. This gives :
  \begin{displaymath}
    - \int_0^T \int_{\vec{x},\theta,\kappa}  f^\varepsilon \, ((\varepsilon\partial_t  + (\vec{\tau}\cdot
      \nabla_{\vec{x}} ) ) \varphi)\,d\kappa
    d\theta  d\vec{x} dt = \varepsilon \int_{\vec{x},\theta,\kappa} f_{0}^\varepsilon \phi_{t=0}\,d\kappa d\theta d\vec{x}.
  \end{displaymath}
  Dividing by $\varepsilon$ and taking the limit $\varepsilon \rightarrow 0$, we get :
  \begin{displaymath}
    - \int_0^T \int_{\vec{x}}  (n^0 \partial_t \varphi +  J^0 \cdot
      \nabla_{\vec{x}} \phi)\, d\vec{x} dt = \int_{\vec{x}} n_{0}^0 \phi_{t=0}\,d\vec{x},
  \end{displaymath}
  where $n_0^0$ is defined by (\ref{ini_cond}). This last equation is exactly the weak  formulation of equation (\ref{continuity}), with initial datum $n_0^0$. This concludes the proof. \endproof

\setcounter{equation}{0}
\section{Equivalence of the two methods}
\label{sec_equivalence}

In this section, we show that both methods lead to the same value of the diffusion coefficient (theorem \ref{thm:Dexplicit}). 

The first step is to show that we can approximate the solution of equation (\ref{eq:solva}) by the
solution of the associated evolution equation. More precisely, in appendix C, we prove the following lemma :

\begin{lemma}
  \label{lem:approx_evo}
  Let $g$ in $H$ satisfying (\ref{solva}) and $u_\infty$ in $D(A)$ be the solution of (\ref{eq:solva}) satisfying (\ref{cancel}). Let $u_0 \in D(A)$ satisfying (\ref{cancel}). Then, the solution $u(t)$ of the evolution problem: 
\begin{equation}
\label{eq:evo_A}
\partial_t u = -A u \, + \,g \, , \quad 
u_{t=0} = u_0, 
\end{equation}
weakly converges to $u_\infty$ in $H$ as $t$ tends to $\infty$. 
\end{lemma}

\medskip
With this lemma we can explicitly calculate the tensor $D$ and prove the theorem
\ref{thm:Dexplicit} :

\medskip

\noindent
{\bf Proof of theorem \ref{thm:Dexplicit}:}  Let $\vec{\chi}(t)$ be the solution of \begin{eqnarray}
\partial_t \vec{\chi} = -A \vec{\chi} \, + \,\vec{\tau}(\theta)\,\frac{M(\kappa)}{2\pi} \, , \quad 
        \vec{\chi}(t=0) = 0. \label{eq_chi(t)}
\end{eqnarray}
Thanks to Lemma \ref{lem:approx_evo}, $\vec{\chi}(t)$ weakly converges to $\vec \chi$ in $H$ when $t \to \infty$. It follows tht :
\begin{equation}
\label{eq:ut_uinf}
\int_{\kappa, \theta} \vec{\chi}(t) \otimes \vec{\tau}\,d\kappa d\theta \stackrel{t
\rightarrow +\infty}{\longrightarrow} \int_{\kappa, \theta} \vec{\chi} \otimes
\vec{\tau}\,d\kappa d\theta.
\end{equation}

Let us consider the first component of $\vec{\chi}(t)$, which we denote by $u(t)$ and the integrals \, $\int_{\kappa, \theta} u(t) \cos \theta \,d\kappa d\theta$ \, and \, $\int_{\kappa, \theta} u(t) \sin \theta \,d\kappa d\theta$.\, Because $u$ satisfies (\ref{eq_chi(t)}), it admits the following representation (see the proof  of Lemma  \ref{lem:approx_evo}):
\begin{displaymath}
u(t) =  \int_0^t T_s\left(\cos \theta\,\frac{M(\kappa)}{2\pi}\right)\,ds,
\end{displaymath}
where $T_t$ is the semi-group generated by $-A$ (see \cite{pazy83:_semig}). With this expression, we evaluate the integral of $u(t)$ against $\cos \theta$ :
  \begin{eqnarray*}
    \int_{\kappa, \theta} u(t) \cos \theta \,d\kappa d\theta &=&  \int_{\kappa, \theta}\int_0^t T_s
    \left (\cos \theta \,\frac{M(\kappa)}{2\pi} \right)\,ds\,
    \cos \theta  \,d\kappa \, d\theta\\
    &=& \int_0^t \int_{\kappa, \theta} \cos \theta \,\frac{M(\kappa)}{2\pi} \, T_s^*
    (\cos \theta)  \,d\kappa \, d\theta \, ds,
  \end{eqnarray*}
where $T^*$ is the adjoint operator of $T$ in $L^2(\theta,\kappa)$ generated by $-{\mathcal A}^*$, where 
\begin{displaymath}
   {\mathcal A}^*(f) = -\kappa \partial_\theta f +\kappa \partial_\kappa f -\alpha^2
\partial_{\kappa^2} f.
\end{displaymath}
Note that we are referring here to the adjoint in the standard $L^2$ sense and not in the weighted space $H$. This is why ${\mathcal A}^*$  does not coincide with $A^*$ defined in Lemma \ref{lem:A*}. The semi-group $T_t^*$ admits a probabilistic representation: for all regular functions $f(\theta,\kappa)$
\begin{displaymath}
    T_t^*(f)(\theta,\kappa) = {\mathbb E}\{f(\theta(t),\kappa(t)) | \theta_0=\theta, \kappa_0=\kappa\},
\end{displaymath}
where $(\kappa(t),\theta(t))$ is the solution of the stochastic differential equation (\ref{angle_scaled}), (\ref{curv_scaled}). Using this representation, we have :
\begin{eqnarray*}
    \int_{\kappa} M(\kappa)\, T_s^* (\cos \theta) \,d\kappa &=&
    \int_{\kappa} M(\kappa)\, {\mathbb E}\{\cos \theta_s | \theta_0=\theta,
    \kappa_0=\kappa\} \,d\kappa \\
    &=& {\mathbb E}\{\cos \theta_s | \theta_0=\theta, \kappa_0=Z\},
\end{eqnarray*}
where $Z$ is a random variable independent of $B_t$ with density $M$. Using lemma \ref{lem:eds}, we have :
\begin{displaymath}
    {\mathbb E}\{\cos \theta_s | \theta_0=\theta, \kappa_0=Z\} = {\mathbb E}\{\cos(\theta + Y_s)\},
\end{displaymath}
with $Y_s$ a Gaussian random variable with zero mean and variance $\beta_s^2$ given by (\ref{beta_explicit}). Then:
\begin{displaymath}
    {\mathbb E}\{\cos(\theta + Y_s)\} = {\mathbb E}\{\cos \theta \, \cos Y_s - \sin \theta \, \sin Y_s \} =
    \cos \theta \, \,  {\mathbb E}\{\cos Y_s \},
\end{displaymath}
because the density of $Y_s$ is even and implies that ${\mathbb E}\{\sin Y_s\}=0$. Finally using (\ref{cos_Z}), we have:
  \begin{displaymath}
    \int_{\kappa} M(\kappa)\, T_s^* (\cos \theta) \,d\kappa = \cos \theta \,\,  {\mathrm e}^{-\frac{\beta_s^2}{2}}.
  \end{displaymath}
Then, the first integral is given by:
  \begin{displaymath}
    \int_{\kappa, \theta} u(t) \cos \theta \, d\kappa \, d\theta = \int_0^t \int_{\theta}
    \frac{\cos \theta}{2\pi} \cos \theta \,  {\mathrm e}^{-\frac{\beta_s^2}{2}} \, d\theta \, ds =
    \int_0^t \frac{1}{2}\,{\mathrm e}^{-\frac{\beta_s^2}{2}} \,ds.
  \end{displaymath}
We can proceed similarly to evaluate the integral of $u(t)$ against $\sin(t)$. This gives:
  \begin{displaymath}
    \int_{\kappa, \theta} u(t) \sin \theta \,d\kappa d\theta = \int_0^t \int_{\theta}
    \frac{\cos \theta \sin \theta }{2\pi}  {\mathbb E}\{\cos Y_s \}\,d\theta ds =0.
  \end{displaymath}
It remains to evaluate the integrals involving the second component of vector
$\vec{\chi}(t)$ which we denote by $v(t)$. By the same method as for $u(t)$, we get :
  \begin{displaymath}
    \int_{\kappa, \theta} v(t) \cos \theta \,d\kappa \, d\theta = 0 \quad \text{ and } \quad
    \int_{\kappa, \theta} v(t) \sin \theta \,d\kappa \, d\theta = \int_0^t \frac{1}{2}\,{\mathrm e}^{-\frac{\beta_s^2}{2}} \,dt.
  \end{displaymath}
Collecting these formulae, we can write:
  \begin{displaymath}
    \int_{\kappa, \theta} \vec{\chi}(t) \otimes \vec{\tau}\,d\kappa d\theta = \frac{\mathcal{D}(t)}{2}\,Id,
  \end{displaymath}
  with $\mathcal{D}(t) = \int_0^t\,{\mathrm e}^{-\alpha^2(-1+u+{\mathrm e}^{-u})}\,du$. Taking the
  limit $t\rightarrow +\infty$ and using equation (\ref{eq:ut_uinf}), shows that (\ref{D=D}) holds true and completes the proof of the theorem. \endproof

\setcounter{equation}{0}
\section{Numerical simulation}
\label{sec_numeric}

We simulate individual trajectories satisfying equation (\ref{pos_scaled})-(\ref{curv_scaled}) with
initial conditions given by (\ref{pos_ini})-(\ref{indep_ini}). If we fix a time step $\Delta t$, using (\ref{angle_explicit}), (\ref{curv_explicit}), we have :
\begin{displaymath}
  \left\{
    \begin{array}{ccl}
      \kappa_{(n+1)\Delta t} &=& \gamma \kappa_{n\Delta t} +  G_{(n+1)} \\
      \theta_{(n+1)\Delta t} &=& \theta_0 \;+\; \kappa_0 - \kappa_{(n+1)\Delta t} + \sqrt{2} \alpha B_{(n+1)\Delta t}.
    \end{array}
  \right.
\end{displaymath}
with $\gamma={\mathrm e}^{-\Delta t}$ and $G_{(n+1)}$ a Gaussian random variable with zero mean
and variance $2\alpha^2\,\left(1-{\mathrm e}^{-2\Delta t}\right)$ independent of $\kappa_{n\Delta t}$. With this
formula, we can simulate recursively the process $(\kappa_{n\Delta t},\theta_{n\Delta
  t})_n$ exactly (in the sense that it has the same law as the exact solution). To
generate the Brownian motion, we just compute the increments $B_{(n+1)\Delta t}
-B_{n\Delta t}$ since they are Gaussian and independent of $B_{n\Delta t}$. On the other
hand, these increments are not independent of $G_{(n+1)}$. Fortunately, we can compute the
covariance matrix of the Gaussian vector $(G_{(n+1)},B_{(n+1)\Delta t}-B_{n\Delta t})$ :
\begin{displaymath}
    \left( \begin{array}{c}
      G_{(n+1)}\\
      B_{(n+1)\Delta t}-B_{n\Delta t}
    \end{array}
  \right) \sim \mathcal{N}(0,C)
\end{displaymath}
where $\mathcal{N}(0,C)$ is a two-dimensional Gaussian vector with zero mean and covariance matrix $C$ given by:
\begin{displaymath}
  C = \left[ \begin{array}{cc}
      2 \alpha^2\,\left(1-{\mathrm e}^{-2\Delta t}\right) & \sqrt{2} \alpha(1-{\mathrm e}^{-\Delta t}) \\
      \sqrt{2} \alpha (1-{\mathrm e}^{-\Delta t}) & \Delta t
    \end{array}\right].
\end{displaymath}
Knowing this covariance matrix, we can simulate the Gaussian vector
$(G_{(n+1)},B_{(n+1)\Delta t}-B_{n\Delta t})$ using the Cholesky method : we generate
$(X_1,X_2)$ a vector of two independent normal law, and take 
$\sqrt{C}(X_1,X_2)^T$ as realization of the Gaussian vector. 

Now for the position $\vec{x}$, since we do not have any explicit expression, we use a
discrete approximation scheme of order $O((\Delta t)^2)$. For example, the first component $x_1$ of $\vec{x}$ is approximated by:
\begin{eqnarray*}
  x_1((n+1)\Delta t) &=& x_1(n\Delta t) + \int_{n\Delta t}^{(n+1)\Delta t}
  \cos \theta(s)\,ds \\
  &\approx& x_1(n\Delta t) + \frac{\Delta t}{2}(\cos \theta(n\Delta
    t)+\cos \theta((n+1)\Delta t)).
\end{eqnarray*}

\begin{figure}
  \centering
  \includegraphics[scale=.8]{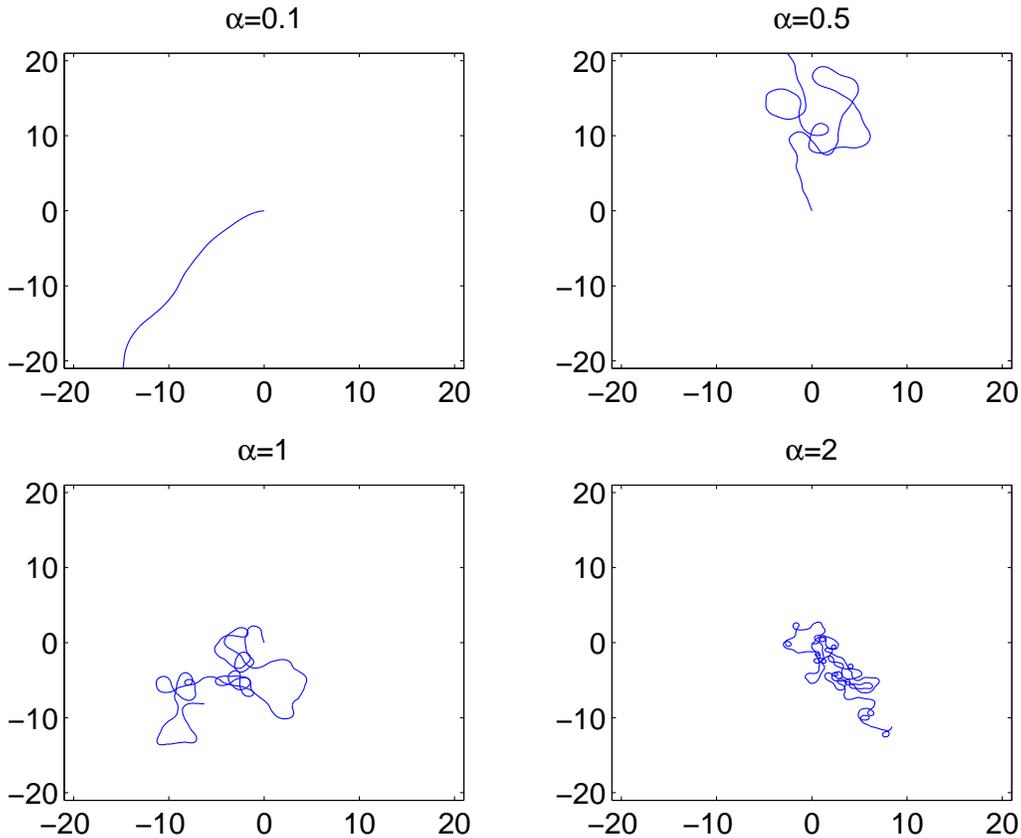}
  \caption{Four trajectories simulated with different value of $\alpha$: $\alpha =0.1$ (top left), $\alpha = 0.5$ (top right), $\alpha = 1.$ (bottom left) and $\alpha = 2.$ (bottom right). The simulation
    is run during 120 time units with a time step $dt=0.05$ time unit}
  \label{fig:simu_traj}
\end{figure}

We present four trajectories obtained with different values of the parameter $\alpha$ in figure \ref{fig:simu_traj}. As the parameter $\alpha$ increases, the excursions towards large positive or negative curvatures become larger. As a consequence, the spinning of the trajectory around itself increases and, from almost a straight line when $\alpha = 0.1$, the trajectory shrinks and looks closer and closer to a wool ball. We can figure out that the diffusion coefficient decreases as $\alpha$ increases which is confirmed by formula (\ref{eq:D}) since the function $-1 + s + \exp(-s) \geq 0$ for $s\geq 0$. Additionally, it is easily seen that ${\mathcal D} \to \infty$ like $\alpha^{-2}$ when $\alpha \to 0$ while ${\mathcal D} \to 0$ when $\alpha \to \infty$. 

\begin{figure}[h]
  \begin{center}
    \begin{tabular}{cc}
      \includegraphics[width=6.85cm,clip=]{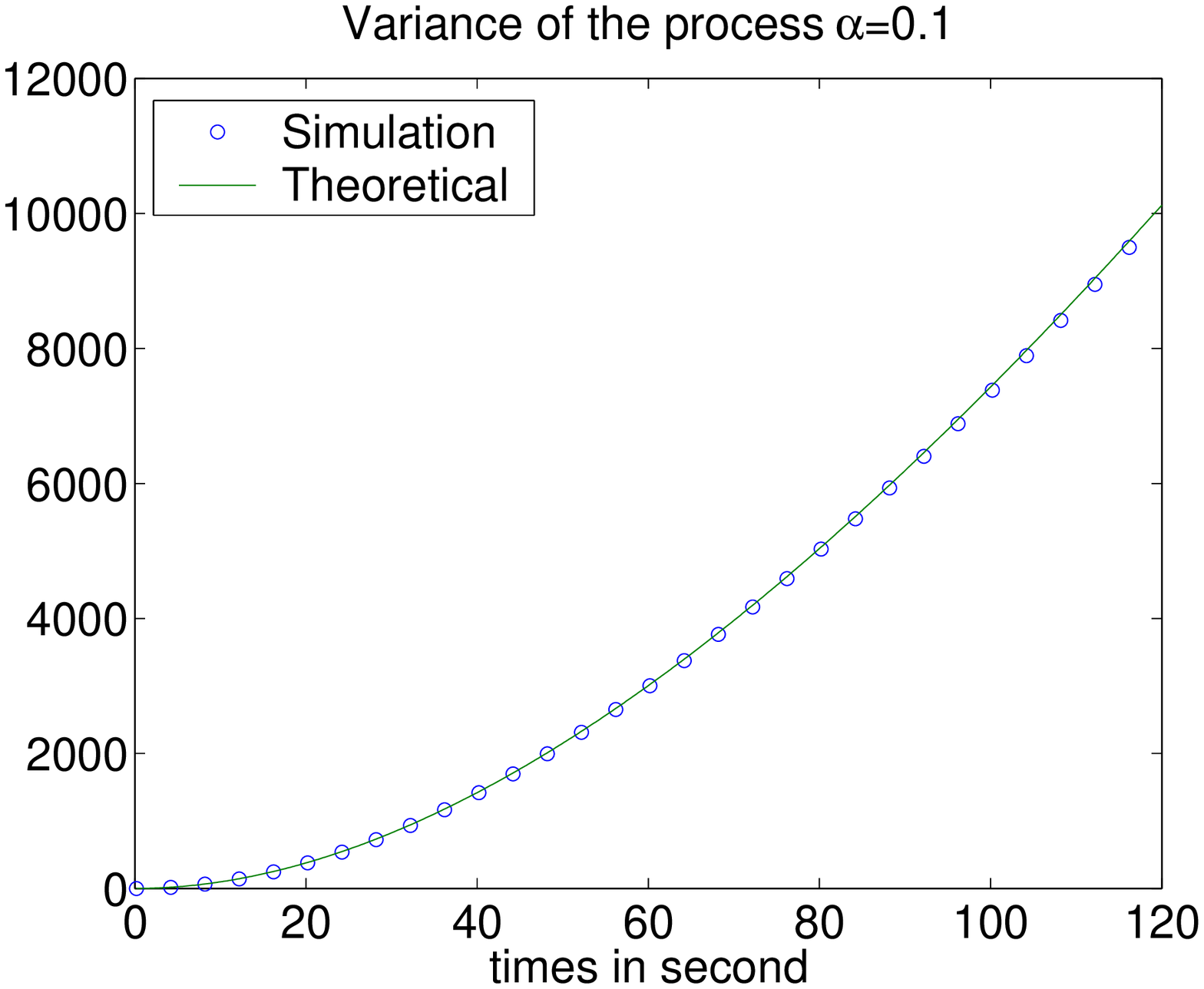}
      &
      \includegraphics[width=6.85cm,clip=]{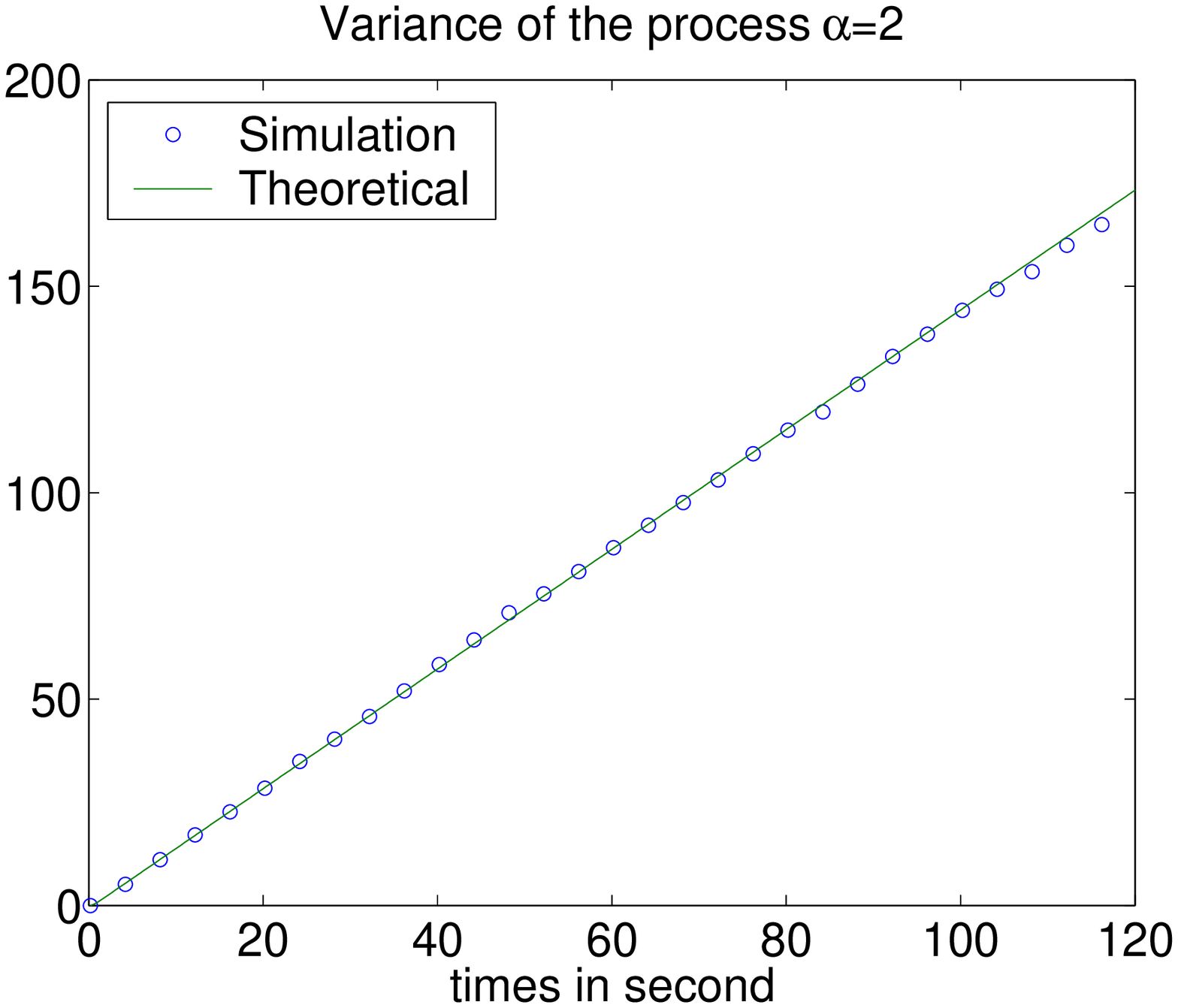} \\
      {\bf (a)}
      &   {\bf (b)}
    \end{tabular}
    \caption{
      \label{fig:simu_var}
      Variance of the process $\vec{x}(t)$: comparison between the numerical simulation (points) and the theoretical prediction (solid line) for two values of $\alpha$ :  $\alpha = 0.1$ (left) and  $\alpha = 2$ (right).}
  \end{center}
\end{figure}

To illustrate theorem \ref{thm:asymp}, we use a Monte-Carlo method to simulate the
variance of the process $\vec{x}$. We simulate $N$ independent trajectories and we compute the variance of the sample at each time step. In figure (\ref{fig:simu_var}), we compare the result obtained with $N=2000$ and the theoretical prediction given by the (\ref{eq:var}). The figure shows an excellent agreement between the computation and the theoretical prediction. Additionally, after an initial transient, the growth of the variance is linear, in accordance with the theoretical result (\ref{eq:D}). 
We can use the slope of the asymptotically linear part of the curve to give a numerical estimate of the diffusion coefficient $\mathcal{D}$. Fro this purpose, we fit a straight line (in the mean-square sense) between times $T/2$ and $T$. We remove the data between $0$ and $T/2$ because the initial transient is not linear and including them would deteriorate the accuracy of the measurement. We compare the slope of the fitted line with the theoretical value (\ref{eq:D}). We report the result of this comparison for two values of $\alpha$ ($\alpha = 0.1$ and $\alpha = 2.$, with $T=1200$ time units in Table \ref{table_1}. The approximation is quite good, with an error comprised between $2$ and $3\%$, which can be attributed to numerical noise and to an unsufficient approximatin of the asymptotic state. 
\begin{table}
\begin{center}
  \begin{tabular}{c|c|c|c}
    & $D$ simulation & $D$ theoretical & relative error \\
    \hline
    $\alpha=0.1$  & $98.5$ & $101$ & $2.5$ \% \\
    \hline
    $\alpha=2$   & $0.708$  & $0.725$ & $2.4$ \%
  \end{tabular}
\caption{Diffusion coefficient: comparison of the numerical estimate obtained by fitting the numerical values with a straight line over the time interval $[T/2,T]$ with the theoretical prediction  (\ref{eq:D}).  $T= 1200$ units of time. }
\label{table_1}
\end{center}
\end{table}
To illustrate the influence of the initial transient, we take $T=120$ time units in the case $\alpha=0.1$ and report the result in Table \ref{table_2}. There, the approximation is quite poor, because the asymptotic state has not yet been reached. 
\begin{table}
\begin{center}
  \begin{tabular}{c|c|c|c}
    & $D$ simulation & $D$ theoretical & relative error \\
    \hline
    $\alpha=0.1$  & $58.8$ & $101$ & $72$ \%
  \end{tabular}
\caption{Same as Table \ref{table_1} but with $T=120$ units of time. The agreement is poor because the asymptotic state is not reached. }
\label{table_2}
\end{center}
\end{table}

\begin{figure}[h!]
  \centering
  \includegraphics[scale=.5]{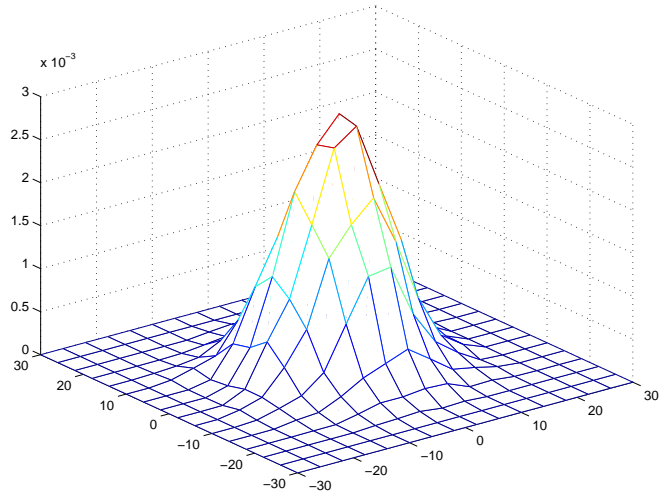}
  \caption{The spatial distribution $n$ for $\alpha=2$ at time $T=30$ time units.}
  \label{fig:simu_distri}
\end{figure}

\begin{figure}[h!]
  \centering
  \includegraphics[scale=.5]{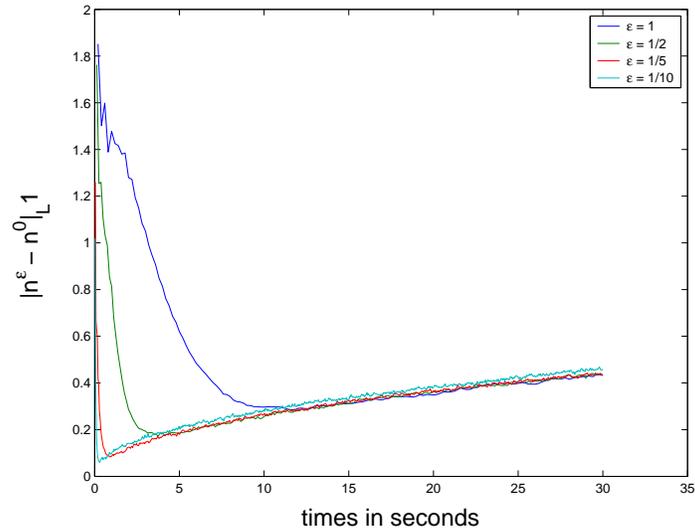}
  \caption{$L^1$ norm of the difference between $n^\varepsilon(t, ) = \int f^\varepsilon( t, \cdot , \theta, \kappa) \, d\theta \, d\kappa$ and its asymptotic limit $n^0(t)$ as a function of time $t$,  with $\alpha=1$ and $N=10^4$ simulation particles.}
  \label{fig:diff_L1}
\end{figure}

In order to illustrate theorem \ref{thm:fasymp}, we plot the spatial density $n(t,\vec{x})$
of the distribution $f(t,\vec{x},\theta,\kappa)$ using a Monte-Carlo algorithm for
$\alpha=2$ and $T=30$ time units on figure \ref{fig:simu_distri}. We see that the density has the Gaussian shape of the solution of a diffusion equation, in accordance with the prediction of the theorem.
To make a more quantitative comparison, we compare it with the asymptotic prediction, i.e. the solution
of the diffusion equation (\ref{continuity}), (\ref{current}), by computing the difference in $L^1$ norm. The results are reported in figure \ref{fig:diff_L1}.
We plot the $L^1$ norm of the difference for $\alpha = 1. $ and for four values of the parameter $\varepsilon$ : $1,\, \frac{1}{2}\,
\frac{1}{5}$ and $\frac{1}{10}$. As expected, the agreement is better as $\varepsilon$ is smaller. However, at large times, all solutions are eventually close to the solution of the diffusion equation. Roughly speaking, the time at which the solution of the diffusion equation starts to be a good approximation of the solution of the kinetic equation scales like $\varepsilon$. This means that, after an initial transient the duration of which may depend on $\varepsilon$, the solution is close to that of the diffusion equation, no matter the value of $\varepsilon$. 

\newpage

\setcounter{equation}{0}
\section{Conclusion}
\label{sec_conclu}

In this paper, the large-scale dynamics of the 'Persistent Turning Walker' (PTW) model of fich behavior has been analyzed. It has been shown, by two different methods, that the large scale limit of this model is of diffusion type, and an explicit formula for the diffusion coefficient has been provided. While the direct analysis of the stochastic trajectories provides a direct route to the value of the diffusion constant, the diffusion approximation of the associated forward Kolmogorov equation, which is of Fokker-Planck type, gives a more systematic way to extend the theory to more complex nonlinear cases. Such a nonlinear situation will be encountered when, in the near future, the nonlinear interactions between the individuals will be introduced within the PTW model. We expect that, in this context, the diffusion approximation methodology will have to be exploited thoroughly to allow access to the large scale behaviour of the system.

\newpage
\setcounter{equation}{0}
\section*{Appendix A: proofs of section \ref{sec_trajectory}}

{\bf Proof of Lemma \ref{lem:eds}:} formula (\ref{curv_explicit}) is standard in the theory of Ornstein-Uhlenbeck processes \cite{oksendal92:_stoch}. To obtain (\ref{angle_explicit}), we integrate $\kappa(t)$ with respect to time:
\begin{displaymath}
    \int_0^t \kappa(s)\,ds = (1-{\mathrm e}^{-t}) \kappa_0 + \sqrt{2} \alpha \int_0^t\int_0^s {\mathrm e}^{-s}{\mathrm e}^{u}\,dB_u ds.
\end{displaymath}
Interchanging the order of integrations and integrating with respect to $s$, we deduce:
\begin{displaymath}
    \int_0^t \kappa(s)\,ds = (1-{\mathrm e}^{-t}) \kappa_0 + \sqrt{2} \alpha \int_{0}^t (1-{\mathrm e}^{-(t-u)})\,dB_u.
\end{displaymath}
Then we develop the integral :
\begin{displaymath}
    \int_0^t \kappa(s)\,ds = (1-{\mathrm e}^{-t}) \kappa_0 + \sqrt{2} \alpha B_t - (\kappa (t)-{\mathrm e}^{-t}\kappa_0). 
\end{displaymath}
This formula can be rewritten:
\begin{displaymath}
    \int_0^t \kappa(s)\,ds = \kappa_0-\kappa (t) + \sqrt{2} \alpha B_t,
\end{displaymath}
which easily leads to (\ref{angle_explicit}). 

We now calculate the mean and the variance of $K_0^t = \int_0^t
  \kappa(s)\,ds$. Since $\kappa(s)$ is of zero mean, its integral $K_0^t$ is also of zero mean: \, ${\mathbb E}\{K_0^t\} =  0$. Now for the variance of $K_0^t$, we can write :
\begin{displaymath}
    \mathrm{Var}\{K_0^t\} ={\mathbb E} \left\{ \left( (1 - {\mathrm e}^{-t}) \kappa_0 - \sqrt{2}\alpha {\mathrm e}^{-t} \int_0^t {\mathrm e}^{s}\,dB_s + \sqrt{2} \alpha B_t \right)^2
    \right\} . 
\end{displaymath}
Using that $\kappa_0$ and $B_s$ are independent, we can develop the square and get:
\begin{displaymath}
    \mathrm{Var}\{K_0^t\} = (1 - {\mathrm e}^{-t})^2 {\mathbb E}\{\kappa_0^2\} + 2\alpha^2
    {\mathbb E} \left\{\left(- {\mathrm e}^{-t}\int_0^t {\mathrm e}^{s}\,dB_s + B_t \right)^2 \right\}.
\end{displaymath}
Let us consider the second term. By Ito's formula, we have
\begin{eqnarray*}
    {\mathbb E} \left\{\left(- {\mathrm e}^{-t}\int_0^t {\mathrm e}^{s}\,dB_s + B_t \right)^2 \right\} &=& {\mathrm e}^{-2t}{\mathbb E}
    \left\{\left(\int_0^t {\mathrm e}^{s}\,dB_s\right)^2 \right\} - 2{\mathrm e}^{-t}{\mathbb E}
    \left\{B_t\int_0^t {\mathrm e}^{s}\,dB_s \right\} + {\mathbb E}
    \left\{B_t^2 \right\}\\
    &=& {\mathrm e}^{-2t}\int_0^t {\mathrm e}^{2s}\,ds - 2{\mathrm e}^{-t}{\mathbb E}
    \left\{B_t({\mathrm e}^t B_t-\int_0^t {\mathrm e}^{s}B_s\,ds \right\} + t,
\end{eqnarray*}
where the Ito correction term is zero due to the fact that $\exp s$ is a deterministic process. We can simplify this expression again since ${\mathbb E}\{B_t B_s\} = \min(t,s)$ and get:
\begin{eqnarray*}
    {\mathbb E} \left\{\left(- {\mathrm e}^{-t}\int_0^t {\mathrm e}^{s}\,dB_s + B_t \right)^2 \right\} &=&
    \frac{1-{\mathrm e}^{-2t}}{2} - 2 {\mathrm e}^{-t}\left({\mathrm e}^t t -\int_0^t {\mathrm e}^s s\,ds\right) + t \\
    &=& \frac{1-{\mathrm e}^{-2t}}{2} - 2 \left(1-{\mathrm e}^{-t}\right) + t.
\end{eqnarray*}
Using also that  ${\mathbb E}\{\kappa_0^2\}=\alpha^2$, the variance of $K_0^t$ is written:
\begin{displaymath}
    \mathrm{Var}\{K_0^t\} = (1 - {\mathrm e}^{-t})^2 \alpha^2 + 2\alpha^2
    \left( \frac{1-{\mathrm e}^{-2t}}{2} - 2 \left(1-{\mathrm e}^{-t}\right) + t \right).
\end{displaymath}
Developing and simplifying the expression, we find (\ref{beta_explicit}), which ends the proof.  \endproof

\newpage
\setcounter{equation}{0}
\section*{Appendix B: proofs of section \ref{sec_asymptotic}}

{\bf Proof of Lemma \ref{lem:A_max}:} Let $u \in D(A)$. Then, $\kappa \partial_\theta u \in V'$ and Lemma A1 of \cite{Degond_AnnENS} shows that the Green formula for functions $u \in V$ such that $\kappa \partial_\theta u \in V'$ is legitimate. Therefore, taking the inner product of $A(u)$ against $u$, we find:
\begin{eqnarray}
    <A(u),u>_H =\int_{\theta, \kappa} \alpha^2 M \left|\partial_\kappa \left( \frac{u}{M} \right) \right|^2\,d\theta d\kappa \geq 0. \label{Auu}
\end{eqnarray}
So, $A$ is a monotone operator in $H$. To show that $A$ is maximal monotone, we  prove that for any $g \in H$, there exists $u \in D(A)$ such that :
\begin{equation}
    \label{eq:max}
    u + Au = g.
\end{equation}
Taking the inner product of (\ref{eq:max}) against a test function $\varphi$ in the space $\mathcal{D}(\Pi_{\theta}\times {\mathbb R}_\kappa)$ of infinitely differentiable and compactly supported functions on $\Pi_{\theta}\times {\mathbb R}_\kappa$ leads to the variational problem :
\begin{equation}
    \label{eq:max_weak}
    \int_{\kappa,\theta} [ u\left(\varphi -
      \kappa\partial_\theta \varphi\right)\,\frac{1}{M} \, + \,M
    \partial_k\left(\frac{u}{M}\right) \partial_k\left(\frac{\varphi}{M}\right) ] \,d\theta
    d\kappa = \int_{\theta, \kappa} g \varphi\,\frac{d\kappa d\theta}{M},
\end{equation}
Again, the same theory as in the appendix A of \cite{Degond_AnnENS} (based the result by J. L. Lions in \cite{lions61:_equat}) applies to prove the existence of a solution to
(\ref{eq:max_weak}) with $u$ in $V$ such that $\kappa \partial_\theta u \in V'$.
From there, it immediately follows that $u \in D(A)$. 

It is immediate to see that any function of the form $u(\theta, \kappa) = C M(\kappa)$ for any constant $C$ belongs to the kernel of $A$. Conversely, suppose that $u \in \mbox{Ker} \,  A$. Then, by (\ref{Auu}), there exists a function $C(\theta) \in L^2(\Pi)$ such that $u(\theta, \kappa) = C(\theta) M(\kappa)$. But again, $A(u)=0$ implies that $\kappa \, 
  \partial_{\theta} C(\theta) \,  M=0$. So $C(\theta)$ is a constant, which proves (\ref{kerA}).
\endproof

\bigskip
\noindent
{\bf Proof of Proposition \ref{ppo:solva}:} the 'only if' part of the theorem is obvious since, using Green's formula (again, obtained by adapting that of appendix B of \cite{Degond_AnnENS}, we have $\int Au \, d\theta \, d \kappa = 0$. 

To prove the 'if' part, we borrow a method from (for instance) \cite{castella06:_diffus}. To find a solution to (\ref{eq:solva}), we look at a perturbed equation :
\begin{equation}
    \label{eq:solva_pertu}
    \lambda u + Au =g,
\end{equation}
with $\lambda >0$. Since $A$ is maximal monotone in $H$ (Lemma \ref{lem:A_max}), eq.
  (\ref{eq:solva_pertu}) admits a solution $u_\lambda$ for all  positive $\lambda$
  (\cite{brezis83:_analy}). To prove the existence of a solution to (\ref{eq:solva}), we want to extract a subsequence, still denoted by $(u_\lambda)$ which converges weakly in $H$. For this purpose, it is enough to show that there exists a bounded subsequence. 

We proceed by contradiction, supposing that the (full) sequence $N_\lambda=\|u_\lambda\|_H \stackrel{\lambda \rightarrow 0}{\rightarrow} + \infty$. 
We define $ U_\lambda= \frac{u_\lambda}{N_\lambda}$.  $U_\lambda$ satisfies $\|U_\lambda\|_H=1$ for all $\lambda$ and
\begin{equation}
    \label{eq:U_ld}
    \lambda U_\lambda + AU_\lambda = \frac{g}{N_\lambda}.
\end{equation}
Since $(U_\lambda)_\lambda$ is a bounded sequence in $H$, we can extract a subsequence (still denoted by $U_\lambda$) such that $U_\lambda \rightharpoonup U$ in $H$ weak as $\lambda \to 0$. Taking the limit $\lambda \rightarrow 0$ in (\ref{eq:U_ld}), gives $A(U) = 0$. If we take the inner product of (\ref{eq:U_ld})
with $U_\lambda$ and then pass to the limit $\lambda \rightarrow 0$, we also find that $U$ belongs to $V$. So Lemma \ref{lem:A_max} applies and gives $U=cM$ with a constant $c \in {\mathbb R}$. Using (\ref{solva}), we also have :
\begin{displaymath}
    <\lambda U_\lambda + AU_\lambda,M>_H = <\frac{g}{N_\lambda},M>_H = 0.
\end{displaymath}
So \, $\int_{\kappa,\theta} U_\lambda \,d\theta d\kappa =0$ \, for all $\lambda$. Taking the limit $\lambda \rightarrow 0$  leads to \, $\int_{\kappa,\theta} U \,d\theta d\kappa = \int_{\kappa,\theta} C M(\kappa) \,d\theta d\kappa = 2 \pi C = 0$, \, which implies $U=0$. This proves:
\begin{equation}
    \label{eq:U_ld_weak}
    U_\lambda \rightharpoonup 0 \qquad \text{ in } H \\ \mbox{ weak }.
\end{equation}
To get a contradiction, we now prove that the convergence is strong.
  
To this aim, we introduce a decomposition of the space $H$ into two orthogonal subspaces. Let $L$ be the closed subspace of $H$ defined by :
\begin{displaymath}
    L = \{ c(\theta)M \, / \; c(\theta) \in L^2(\Pi_{\theta}) \},
\end{displaymath}
with $M$ defined by (\ref{eq:M}). So $H= L \stackrel{\perp}{\oplus} L^{\perp}$. We also define the orthogonal projector $P$ of $H$ onto $L$ such that $Pf =  \left(\int_{\kappa} f(\kappa,\theta)\,d\kappa\right) M$. Using this projection, we decompose the sequence $(U_\lambda)_\lambda$ as follows:
\begin{equation}
    \label{eq:dec}
    U_\lambda = c_\lambda(\theta)M \, + \, v_\lambda,
\end{equation}
with $v_\lambda \in L^{\perp}$, i.e. $\int_\kappa v_\lambda \,d\kappa = 0$. To demonstrate that $U_\lambda \stackrel{\lambda \rightarrow 0}{\longrightarrow}
  0$ in $H$ strongly, we first demonstrate that $v_\lambda \stackrel{\lambda \rightarrow 0}{\longrightarrow} 0$ in $H$ strongly.
  
Taking the inner product of the equation satisfied by $U_\lambda$ (\ref{eq:U_ld}) with $U_\lambda$ gives :
\begin{displaymath}
    \lambda \|U_\lambda\|_H^2 + \int_{\theta,\kappa} M \left|\partial_\kappa
      \frac{U_\lambda}{M}\right|^2\,d\theta d\kappa = \frac{1}{N_\lambda}\,<g,U_\lambda>_H.
\end{displaymath}
Since $\partial_\kappa \frac{U_\lambda}{M} = \partial_\kappa \frac{v_\lambda}{M}$ and $\|U_\lambda\|_H=1$, we get by taking the limit $\lambda \rightarrow 0$ :
\begin{equation}
    \label{eq:dv_ld}
    \int_{\theta,\kappa} M \left|\partial_\kappa
      \frac{v_\lambda}{M}\right|^2\,d\theta d\kappa \stackrel{\lambda\rightarrow
      0}{\longrightarrow } 0.
\end{equation}
Now Gross inequality \cite{gross92:_logar_sobol} gives, for any $v \in V$: 
\begin{equation}
    \label{eq:Gross}
    \int_{\mathbb R} \left| \partial_{\kappa} \left( \frac{ f}{M} \right) \right|^2 M\,d\kappa + \left( \int_{\mathbb R} f \,d\kappa \right)^2 \geq
    \int_{\mathbb R} |f|^2 \, \frac{d\kappa}{M} . 
\end{equation}
Then, since $\int_\kappa v_\lambda \,d\kappa =0$, we deduce:
\begin{displaymath}
    \int_{\mathbb R} \left| \partial_\kappa \frac{v_\lambda}{M} \right|^2 M\,d\kappa  \geq
    \int_{\mathbb R} \frac{|v_\lambda|^2}{M}\,d\kappa.
\end{displaymath}
Integrating this inequality with respect to $\theta$ and using (\ref{eq:dv_ld}), we find:
\begin{eqnarray}
    \|v_\lambda\|_H \stackrel{\lambda\rightarrow 0}{\longrightarrow } 0, \quad \mbox{ in } H \mbox{ strong}. \label{v_to_0}
\end{eqnarray}

To prove the convergence of $c_\lambda$, we define the bounded operator $T$:  $H \to L^2(\Pi_{\theta})$ such that $Tf =  \int_\kappa \kappa f\,d\kappa$. Having $T$ acting on (\ref{eq:U_ld}) and taking the limit $\lambda \rightarrow 0$, leads to:
\begin{equation}
    \label{eq:T_A}
    T \, A \, U_\lambda  \stackrel{\lambda\rightarrow 0}{\longrightarrow } 0 \quad \mbox{ in } L^2(\Pi_{\theta}) \mbox{ strong}.
\end{equation}
If we develop the left-hand side, we find:
\begin{eqnarray*}
    T \, A \, U_\lambda  &=&  \int_\kappa \kappa^2 \partial_\theta U_\lambda \,d\kappa -
    \int_\kappa \left[ \kappa \partial_\kappa(\kappa U_\lambda) -\alpha^2 \kappa \partial_{\kappa^2}
      U_\lambda\right] \,d\kappa\\
    &=& \int_\kappa \kappa^2 \partial_\theta U_\lambda \,d\kappa +
    \int_\kappa \kappa U_\lambda  \,d\kappa.
\end{eqnarray*}
But using the decomposition $U_\lambda = c_\lambda M + v_\lambda$ (\ref{eq:dec}), we have:
\begin{displaymath}
    \left\| \int_\kappa \kappa U_\lambda  \,d\kappa\right\|_{L^2(\theta)} 
    = \left\| \int_\kappa \kappa v_\lambda  \,d\kappa\right\|_{L^2(\theta)} \stackrel{\lambda\rightarrow 0}{\longrightarrow } 0.
\end{displaymath}
So, (\ref{eq:T_A}) leads to :
  \begin{equation}
    \label{eq:Tineg}
    \int_\kappa \kappa^2 \partial_\theta U_\lambda \,d\kappa \stackrel{\lambda\rightarrow 0}{\longrightarrow } 0 \quad \mbox{ in } L^2(\Pi_{\theta}) \mbox{ strong}.
  \end{equation}
If we define $h_\lambda(\theta)= \int_\kappa \kappa^2  U_\lambda \,d\kappa$, (\ref{eq:Tineg}) is equivalent to saying that  $\|\partial_\theta h_\lambda \|_{L^2(\theta)} \stackrel{\lambda\rightarrow
    0}{\longrightarrow } 0$. Using the Poincare-Wirtinger inequality \cite{brezis83:_analy}, there exists a constant $C_0$ such that:
\begin{equation}
    \label{eq:Wirt}
    \|h_\lambda-\bar{h}_\lambda\|_{L^2(\theta)} \leq C_0 \|\partial_\theta h_\lambda \|_{L^2(\theta)},
\end{equation}
with $\bar{h}_\lambda = \frac{1}{2\pi}\,\int_0^{2\pi} h_\lambda(\theta)\,d\theta$. Then, we  develop $\bar{h}_\lambda$. We get:
\begin{displaymath}
    \bar{h}_\lambda = \frac{1}{2\pi}\,\int_0^{2\pi} \!\!\!\!\! \int_\kappa \kappa^2  U_\lambda \,d\kappa
    d\theta= <U_\lambda, M\kappa^2>_H  \;\stackrel{\lambda\rightarrow 0}{\longrightarrow }
    0 \quad \mbox{ in }{\mathbb R},
\end{displaymath}
since $U_\lambda$ converges weakly to zero (see (\ref{eq:U_ld_weak})). So,  (\ref{eq:Wirt})  leads to $h_\lambda \stackrel{\lambda\rightarrow 0}{\longrightarrow } 0$ in $L^2(\Pi_{\theta})$ strong. If we develop $h_\lambda$ we find:
\begin{displaymath}
    h_\lambda(\theta) =  \int_\kappa \kappa^2 (c_\lambda(\theta) M + v_\lambda) \,d\kappa =\alpha^2 c_\lambda(\theta) + \int_\kappa \kappa^2  v_\lambda \,d\kappa.
\end{displaymath}
Now, $\int_\kappa \kappa^2  v_\lambda \,d\kappa$ converges to zero in $L^2(\theta)$ strong because of (\ref{v_to_0}) and we finally have :
\begin{displaymath}
    c_\lambda(\theta)\stackrel{\lambda\rightarrow 0}{\longrightarrow } 0 \; \text{ in } L^2(\theta) \mbox{ strong}.
\end{displaymath}

\medskip
Using the convergence of $c_\lambda$ and $v_\lambda$, we can now prove the strong convergence of  $U_\lambda$ to $0$ in $H$:
\begin{displaymath}
    \|U_\lambda\|_H^2 = \|c_\lambda M\|_H^2 + \|v_\lambda\|_H^2 = \|c_\lambda\|_{L^2(\theta)}^2 + \|v_\lambda\|_H^2 \stackrel{\lambda\rightarrow 0}{\longrightarrow } 0,
\end{displaymath}
which contradicts the fact that $U_\lambda$ has unit norm in $H$. This shows that there exists a bounded subsequence in the sequence $u_\lambda$. In fact, since the same proof can be applied to any subsequence, this shows that the whole sequence $u_\lambda$ is bounded, but this is useless for our purpose. 

We conclude the proof of Proposition \ref{ppo:solva} as follows: there exists a
subsequence $u_\lambda$ and a function $u$ in H such that $u_\lambda \rightharpoonup u$ in $H$ weak. Taking the limit of (\ref{eq:solva_pertu}) as $\lambda \to 0$, we deduce that $Au = g$ in the sense of distributions. However, since $g \in H$, eq. $Au=g$ also holds in $H$. Moreover if we take the inner product of (\ref{eq:solva_pertu}) with $u_\lambda$ and pass to the limit $\lambda \rightarrow 0$, we find that $u$ belongs to $V$. So $u$ belongs to $D(A)$, which ends the proof of the 'if' part of the statement. 

Finally, to prove uniqueness, we just remark that, two solutions of (\ref{eq:solva}) differ from an element of the kernel of $A$ and we apply (\ref{kerA}). This ends the proof. 
\endproof

\newpage
\setcounter{equation}{0}
\section*{Appendix C: proofs of section \ref{sec_equivalence}}

{\bf  Proof of Lemma \ref{lem:approx_evo}:} the proof borrows some ideas from \cite{Desvillettes_Dolbeault}, but is simpler, due to the linear character of the problem. The difficulty is getting some compactness in time. Here, instead of considering time translates of the solution as in \cite{Desvillettes_Dolbeault}, we will consider time integrals over a fixed interval length $\Delta t$. 

Since operator $A$ is maximal monotone on $H$ (see Lemma \ref{lem:A_max}), operator $-A$ generates a semi-group of contractions $T_t$ on $H$. Moreover the solution of (\ref{eq:evo_A}) is given by:
\begin{displaymath}
    u(t) = T_t(u_0) +  \int_0^t T_s(g)\,ds.
\end{displaymath}
We define $f(t)=u(t)-u_\infty$ which satisfies :
\begin{equation}
    \label{eq:ff}
        \partial_t f = -A f , \quad 
        f_{t=0} = f_0, 
\end{equation}
with $f_0=u_0-u_\infty$ and $\int_\kappa f_0(\kappa)\,d\kappa=0$. To prove the weak
convergence of $u(t)$ to $u_\infty$, we have to prove that $f(t)$ converges to zero weakly in $H$.

To this aim, we make an orthogonal decomposition of $f(t)$ as in the proof of
Proposition \ref{ppo:solva}: $f(t) = c(t) M + v(t)$, with $c(t) \in L^2(\Pi_{\theta})$, $v(t) \in H$ and $\int_\kappa v(t)\,d\kappa = 0$.
Taking the inner product of (\ref{eq:ff}) with $f$, we get :
\begin{displaymath}
    \frac{1}{2}\,\partial_t \|f\|_H^2 = -\int_{\kappa, \theta} \alpha^2 M \left[ \partial_\kappa
      \left( \frac{f}{M} \right) \right]^2\,d\kappa d\theta.
\end{displaymath}
Using the decomposition of $f(t)$ and noticing that $\partial_\kappa \left(
    \frac{f}{M} \right)= \partial_\kappa \left( \frac{v}{M} \right)$, this equality becomes:
\begin{eqnarray}
    \frac{1}{2}\,\partial_t \left( \|c(t)\|_{L^2}^2 + \|v(t)\|_H^2 \right) = -\int_{\kappa,
      \theta} \alpha^2 M \left[ \partial_\kappa \left( \frac{v(t)}{M} \right) \right]^2\,d\kappa d\theta. \label{L2_C}
\end{eqnarray}
If we apply the Gross inequality (\ref{eq:Gross}), we get:
  \begin{displaymath}
    \frac{1}{2}\,\partial_t \left( \|c(t)\|_{L^2}^2 + \|v(t)\|_H^2 \right) \leq -\|v(t)\|_H^2.
  \end{displaymath}
Since $c(t)$ is bounded by $\|f_0\|_H^2$, by integrating with respect to time, we have :
  \begin{displaymath}
    \frac{1}{2} \|v(t)\|_H^2 \leq - \int_0^t \|v(s)\|_H^2\,ds + C.
  \end{displaymath}
Using the Gronwall lemma, we deduce that $v(t)$ decays exponentially fast to zero strongly in $H$:
\begin{displaymath}
    v(t) \; \stackrel{t \rightarrow +\infty}{\longrightarrow} \; 0 \; \text{ in } H \mbox{ strong}.
\end{displaymath}

It remains to prove the convergence of $c(t)$ to zero. We integrate (\ref{eq:ff}) with respect to $\kappa$. This gives, using that $\int_\kappa
M(\kappa)\,d\kappa=1$ and $\int_\kappa v(t)\,d\kappa=0$ :
\begin{equation}
    \label{eq:c_t}
    \partial_t c(t) = \partial_\theta \int_\kappa \kappa v(t)\,d\kappa.
\end{equation}
Now if we pre-multiply by $\kappa$ before integrating with respect to $\kappa$, we obtain :
\begin{equation}
    \label{eq:c_theta}
    \partial_t \int_\kappa \kappa v(t)\,d\kappa = \alpha^2 \partial_\theta c(t) +
    \partial_\theta \int_\kappa \kappa^2 v(t) \,d\kappa -\int_\kappa \kappa v(t)\,d\kappa.
\end{equation}
We fix a time interval $\Delta t$ and integrate (\ref{eq:c_theta}) over this time interval. This leads to:
\begin{eqnarray*}
    \int_\kappa \kappa (v(t+\Delta t)-v(t))\,d\kappa &=& \alpha^2 \partial_\theta
    \int_t^{t+\Delta t} \!\!\!\! c(s)\,ds +
    \partial_\theta \int_\kappa  \kappa^2 \int_t^{t+\Delta t} \!\!\!\! v(s)\,ds \,d\kappa
    \\
    && \qquad -\int_\kappa \kappa \int_t^{t+\Delta t}\!\!\!\! v(s)\,ds \,d\kappa.
\end{eqnarray*}
Since $v(t)$ converges to zero in $H$, we have, in the sense of distributions:
\begin{eqnarray}
    \alpha^2 \partial_\theta \int_t^{t+\Delta t} c(s)\,ds \stackrel{t \rightarrow
      +\infty}{\rightharpoonup} 0. \label{part_theta_int_t}
\end{eqnarray}
Since $c$ belongs to $L^\infty((0,\infty)_t,L^2(\Pi_\theta))$ (see (\ref{L2_C}), we have $\int_t^{t+\Delta t} c(s)\,ds \in L^\infty((0,\infty)_t,L^2(\Pi_\theta))$. So there exists a subsequence such that $\int_t^{t+\Delta t} c(s)\,ds$ is weakly convergent in $L^2(\Pi_\theta)$. Actually, (\ref{part_theta_int_t}) implies that there exists a constant function with respect to $\theta$, depending on $\Delta t$
and denoted by $L(\Delta t)$ such that 
\begin{displaymath}
    \int_t^{t+\Delta t} c(s)\,ds \stackrel{t \rightarrow
      +\infty}{\rightharpoonup} L(\Delta t).
\end{displaymath}
To deduce the convergence of $c(t)$, we have to control the derivative of $c(t)$ in
time. For this purpose, we rewrite :  
\begin{eqnarray*}
    \int_t^{t+\Delta t}\!\!\!\! c(s)\,ds &=& \int_0^{\Delta t} \left( c(t) + \int_0^s \partial_t c(t+z)\,dz
    \right) \,ds \\
    &=& \Delta t\, c(t) + \int_0^{\Delta t} \int_0^s \partial_\theta
    \int_\kappa \kappa v(t+z)\,d\kappa \,dz ds.
\end{eqnarray*}
Using again the convergence of $v(t)$ to zero, we find :
\begin{displaymath}
    \Delta t\, c(t) \stackrel{t \rightarrow +\infty}{\rightharpoonup} L(\Delta t),
\end{displaymath}
or defining the constant $C=\frac{L(\Delta t)}{\Delta t}$, we have $c(t) \stackrel{t \rightarrow +\infty}{\rightharpoonup} C$ in $L^2(\Pi_\theta)$ weak.
  
To complete the proof, it remains to prove that $C$ is equal to zero. Now, since eq. (\ref{eq:ff}) is mass preserving i.e.:
\begin{displaymath}
    \partial_t \int_{\kappa, \theta} f(t)\,d\kappa d\theta = -\int_{\kappa, \theta}
    Af(t)\,d\kappa d\theta =0,
\end{displaymath}
we have $ \int_{\kappa, \theta} f(t)\,d\kappa d\theta = \int_{\kappa, \theta}
  f(0)\,d\kappa d\theta =0$.  Also :
  \begin{displaymath}
    \int_{\kappa, \theta} f(t)\,d\kappa d\theta = \int_{\kappa, \theta} (c(t)M +
    v(t))\,d\kappa d\theta \stackrel{t \rightarrow +\infty}{\rightharpoonup} \int_{\theta} C\,d\theta =
    2\pi C.
  \end{displaymath}
  So $C=0$. This proves $f(t) \stackrel{t \rightarrow +\infty}{\rightharpoonup} 0$ in $H$ weak and completes the proof. \endproof


\end{document}